# A Review on Superplastic Forming of Ti-6Al-4V Alloy

Sai Pratyush Akula, Akhil Agnihotri, Amit Kumar Gupta

Department of Mechanical Engineering, BITS Pilani, Hyderabad Campus, Telangana, India – 500078

Keywords: Superplastic forming, Superplasticity, Ti-6Al-4V, Microstructure, Multistep forming dies, etc

## ABSTRACT

This paper presents a review of a different kind of forming process for Ti-6Al-4V known as superplastic forming which has been employed on this alloy to manufacture parts of complex shapes and geometries. This review paper serves as a means to give an outline on all the major works that have been carried out on this front in the past three decades. It covers various aspects related to superplastic forming of titanium alloys like the experimental setups which include manufacturing dies and their modifications to maintain the alloy thickness uniformity after forming. A detailed study of the process parameters has also been done to note the most important physical conditions required for successful forming. This is followed by the influence of microstructure, modern applications of SPF of different titanium alloys and is concluded with an insight for future work and progress in this field.

## 1 INTRODUCTION

It refers to the property of metals to achieve high elongation of over 1000% of without necking, usually at elevated temperatures (about 900°C for Ti-6Al-4V) under a constant, controlled strain rate. The optimum strain rate varies with the superplastic material and this is attributed to the viscous behavior shown by metals and alloys with refined and stable microstructure at elevated temperatures.

Superplastic forming of sheet metal has been used to produce complex patterns and integrated structures that are light-weight and stronger than the assembled components. This results in reduction in the total number of parts and often gives post-processing benefits such as enhanced product performance and cost reduction [7]. Titanium has high specific strength and good cold-formability resulting in applications in manufacturing of aircraft and automobile parts such as compressor blades and turbine discs. However, the technique for improving its formability at elevated temperatures is still a challenge at present. Over the last few decades, a growing interest in superplasticity of Titanium has been seen which can be seen in *Figure 1* and *Figure 2*.

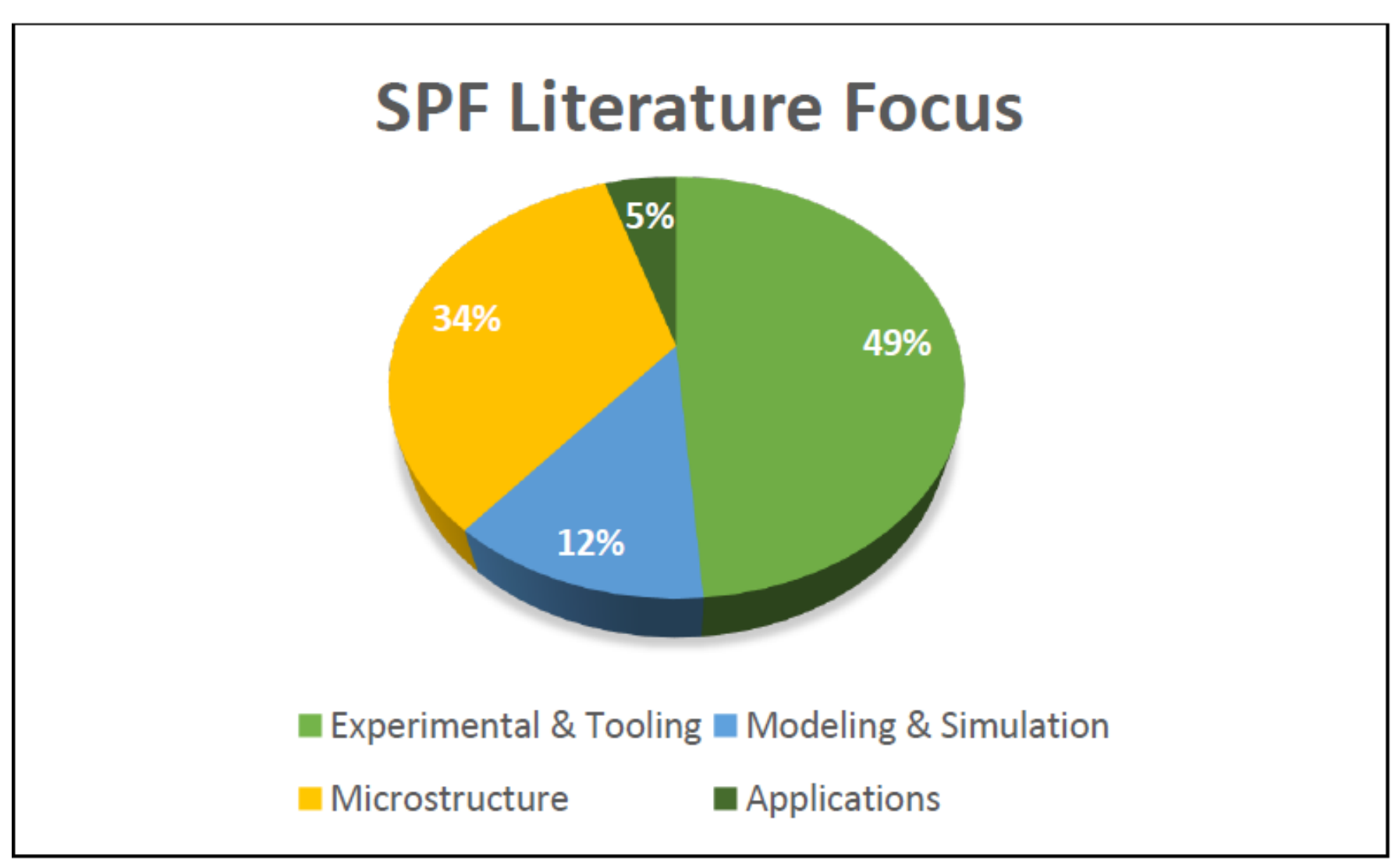

*Figure 1: SPF literature focus in the papers*

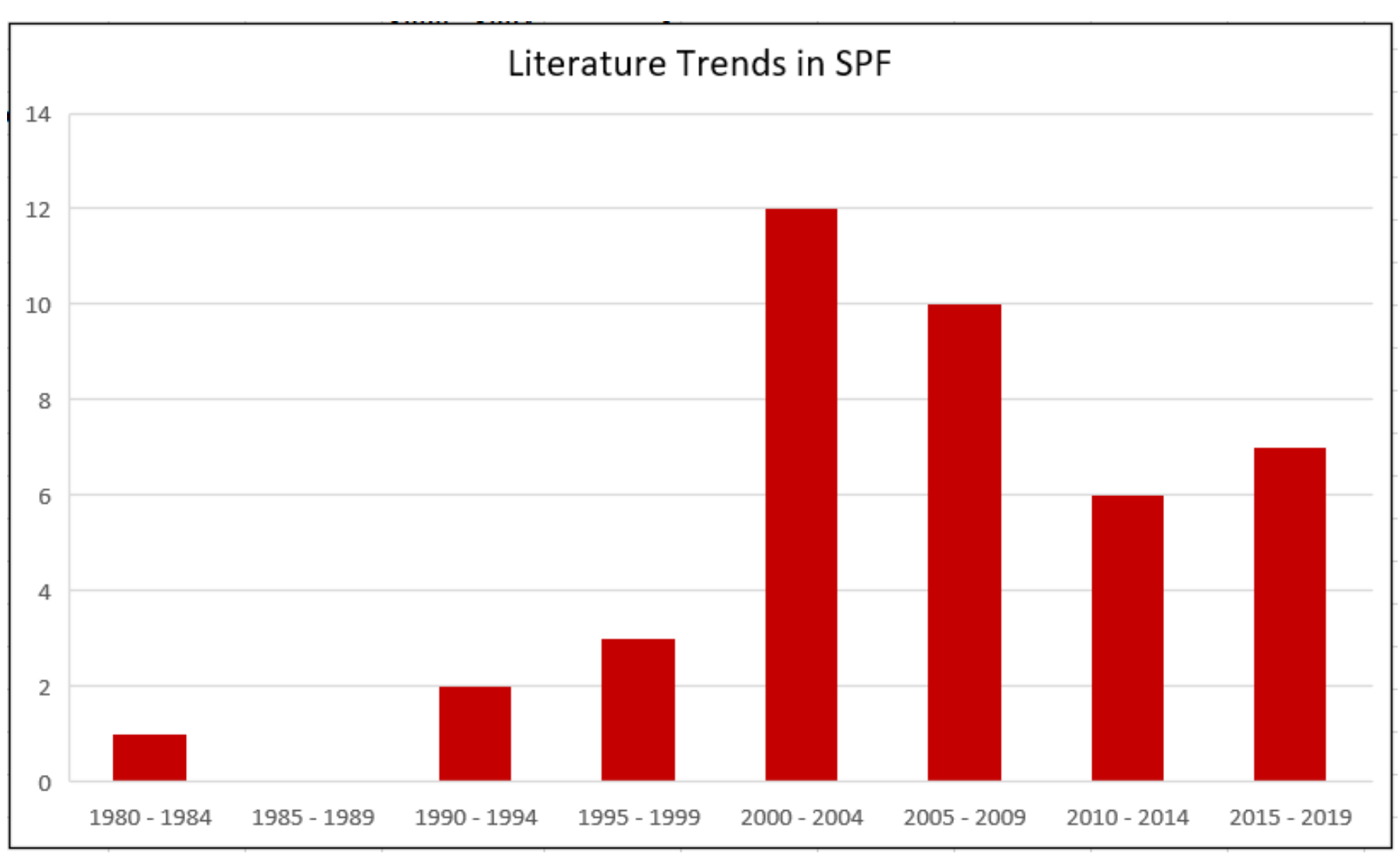

*Figure 2: Literature trends in SPF since 1980*

Discontinuous metal matrix composites (MMCs) possess high strength but low ductility, and are difficult to form with traditional techniques like as rolling, forging or bending [23]. Moreover, due to high hardness, there are only limited number of practically plausible machining techniques. Powder metallurgy does not provide for high dimensional tolerances while casting is only possible in cases where the liquid matrix is non-reactive with the reinforcing phase and the environment. In such cases, superplastic forming is desired since it allows for large strains with limited necking and cavitation.

During SPF, differential straining across the sheet takes place causing large amounts of variation in thickness, thus reducing the sheet's applicability and increasing post-processing costs [7]. This straining occurs in two parts. First, is during free forming (when the sheet has not come in contact with the die), which can be controlled if the alloy exhibits high strain rate sensitivities of flow stress. Second, when the alloy comes in contact with the die although, differential straining is locally controlled, other parts continue to experience this phenomenon. Thus, the last part of the alloy to come in contact with the die has experienced the most strain and hence, has the least thickness. This can be controlled by ensuring that the alloy, more or less, comes into contact with the die in a uniform manner and that the experimental parameters such as load and forming temperature are specific to the local strain gradients.

Bellows expansion joints find application in areas involving expansions and contractions of pipes due to temperature changes, mechanical vibrations, non-uniform support structure, etc. [24]. Superplastically formed using gas pressure and compressive axial load, Titanium alloy Ti-6Al-4V is an ideal candidate for U-type bellows expansion joints as it allows for more than 1000% elongation. The entire process was divided into tube blank fabrication and the actual SPF process.

## 2 Applications of superplastic forming

Titanium alloys are widely used in biomedical applications like cranioplasty due to its biocompatibility along with excellent mechanical features and favors the osseointegration with bones. This makes the alloy suitable for the making of prostheses. Usually, prostheses are machines which are usually expensive and consume a lot of time. Due to this, superplastic forming has found its application in the making of prostheses using titanium alloys [20]. Ambrogio et al., 2018; has work related to the testing and manufacturing of cranial prosthesis using superplastic forming. The cranial prostheses developed by Ambrogio et al., 2018 can be seen in *Figure 3* and the die setup can be seen in *Figure 4.*

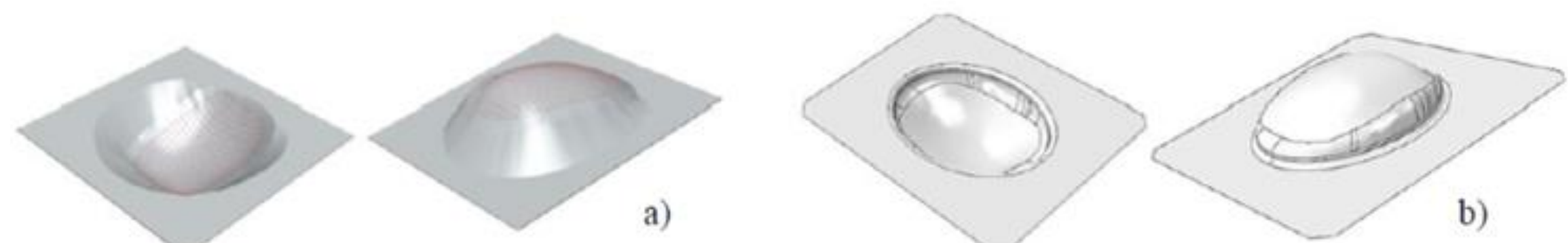

*Figure 3: Prostheses a) made from incremental forming b) made from superplastic forming*

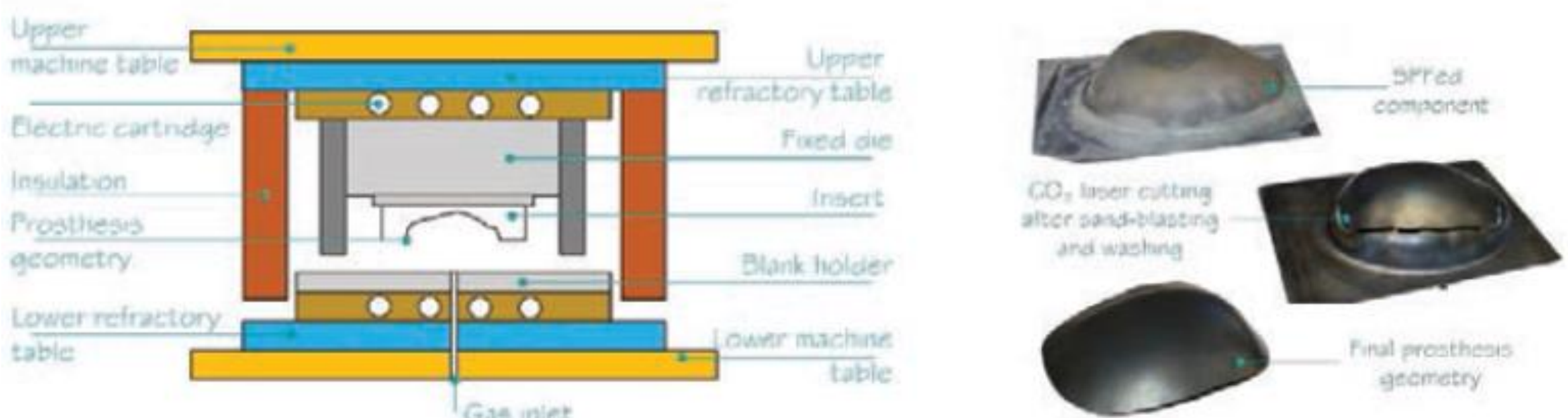

*Figure 4: a) Die setup for SPF of cranial prostheses b) Formed part*

Sorgente, Palumbo, Piccininni, Guglielmi & Aksenov, 2018; have performed a study on the thickness distribution on the prostheses made with the help of finite element simulations and free inflation tests at different pressures. The results of this study can be found in *Table 1.*

*Table 1: Strain rate vs thickness of the component*

| **Target strain rate, $s^{-1}$** | **Minimum thickness (path), mm** | **Minimum thickness (blank), mm** | **Forming time, s** |
|---|---|---|---|
| $2 \times 10^{-5}$ | 0.669 | 0.635 | 61900 |
| $4 \times 10^{-5}$ | 0.667 | 0.647 | 33600 |
| $8 \times 10^{-5}$ | 0.685 | 0.664 | 16175 |
| $1 \times 10^{-4}$ | 0.682 | 0.665 | 12625 |
| $2 \times 10^{-4}$ | 0.653 | 0.653 | 7300 |
| $4 \times 10^{-4}$ | 0.644 | 0.643 | 3775 |
| $1 \times 10^{-3}$ | 0.615 | 0.614 | 1425 |

Superplastic forming has a huge application in the area of aerospace products. Aircraft manufacturing giants like Boeing use SPF to make a few components for their aircrafts [4]. Hefti, 2010; in brief explains the methods and problems related to SPF of Boeing 757 aircraft part. According to the author, superplastic forming along with diffusion bonding finds its application in the fabrication of heat shield assembly and a tailcone for a twin-aisle aircraft (*Figure 5 and 6)*. These said part are superplastically formed using fine grain Ti-6Al-4V at a temperature of 750 - 775°C.

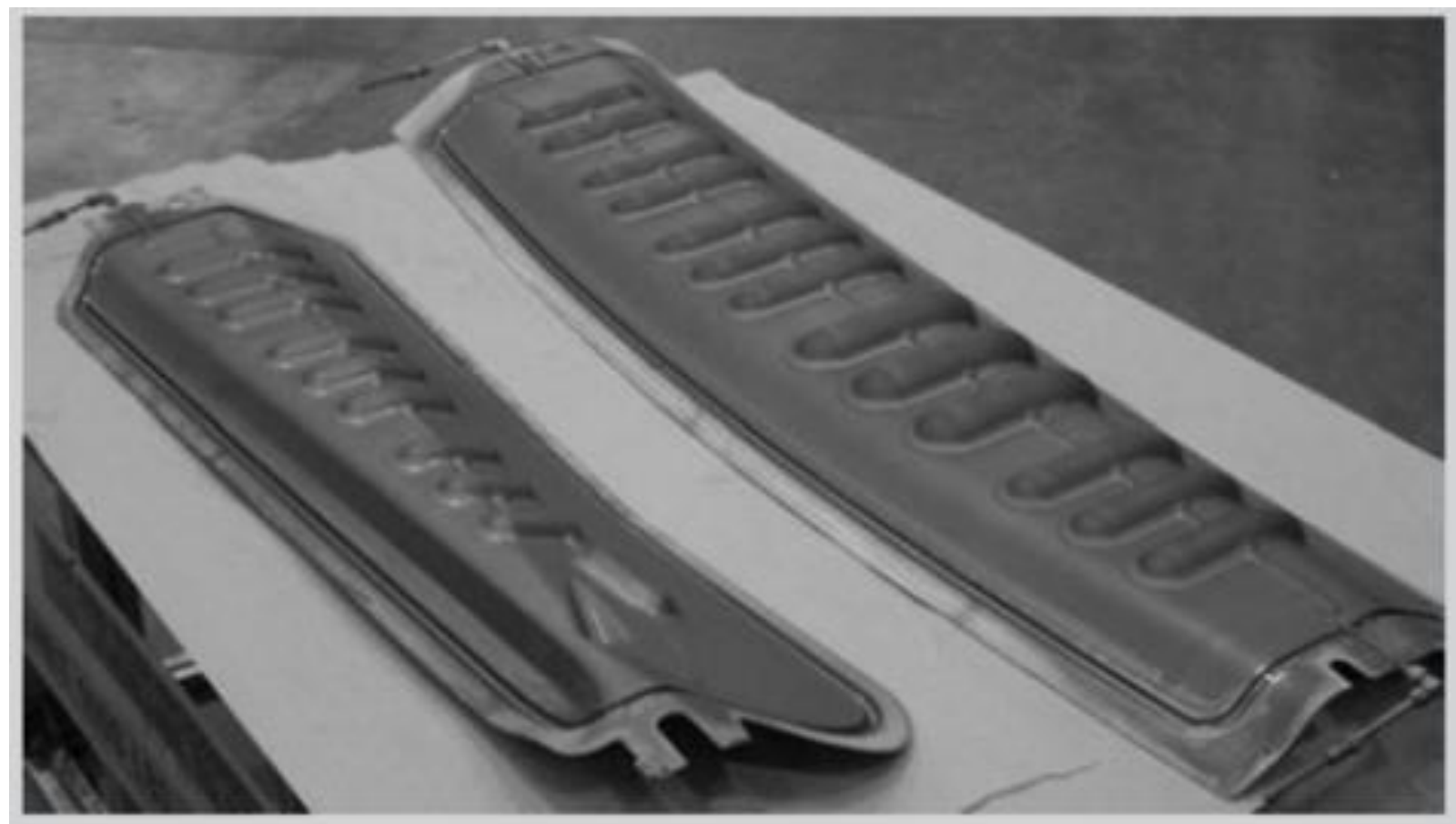

*Figure 5: Twin-aisle aircraft heat shield*

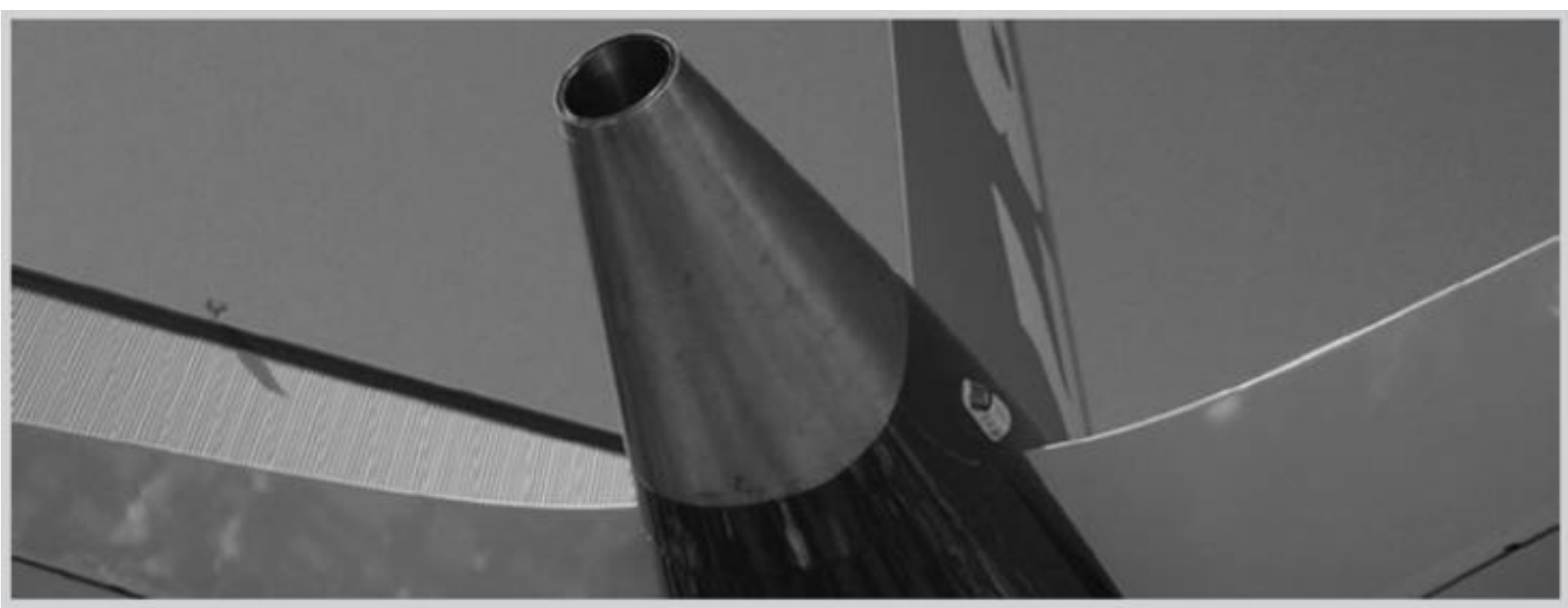

*Figure 6: Tail cone of a twin-aisle aircraft*

The above section has a brief description of the utility of superplastic forming in industries. The methods and the setup for this process will be discussed in the next section. It will contain brief explanations about various dies that can be used, how the process is carried out and the influence of various process parameters in manufacturing a component using superplastic forming.

# 3 Experimental Setup

## 3.1 Dies

### 3.1.1 Single Step Dies

1. Ceramic dies provide superior formability of Ti-6Al-4V due to their adjustable coefficient of thermal expansion [25]. With this regard, a ceramic die was prepared by sintering the green of the $ZrO_2$-$TiO_2$ with the polyvinyl alcohol (PVA) in the atmosphere. Yttria semi-stabilized zirconia (3Y-$ZrO_2$) was used as the basic ceramic die material and the addition of $TiO_2$ to $ZrO_2$ allowed for adjustable coefficient of linear thermal expansion (CTE) and prevented chemical reactions between the die and the Ti-6Al-4V alloy. It is imperative to determine the CTE of the die relative to the alloy for dimensional accuracy in blow-forming. To measure the CTE of the $ZrO_2$-$TiO_2$ die, a 10 x 50 mm cylinder was manufactured in four steps [26]. Firstly, $ZrO_2$ and $TiO_2$ powders along with 25% PVA were mixed in an agitator for 1 hour at 120°C. Then, the green of $ZrO_2$ and $TiO_2$ were pressed into the cylinder mould at 100°C followed by cooling to room temperature over a period of 3 hours after which de-binding and pre-sintering were performed over a range of 25-1000°C for 24 hours. This was concluded by sintering the ceramic cylinder at 1500°C for 1 hour. The schematic for this process is shown in *Figure 7* while *Figure 8* shows the die setup for SPF forming along with its parts labelled.

2.

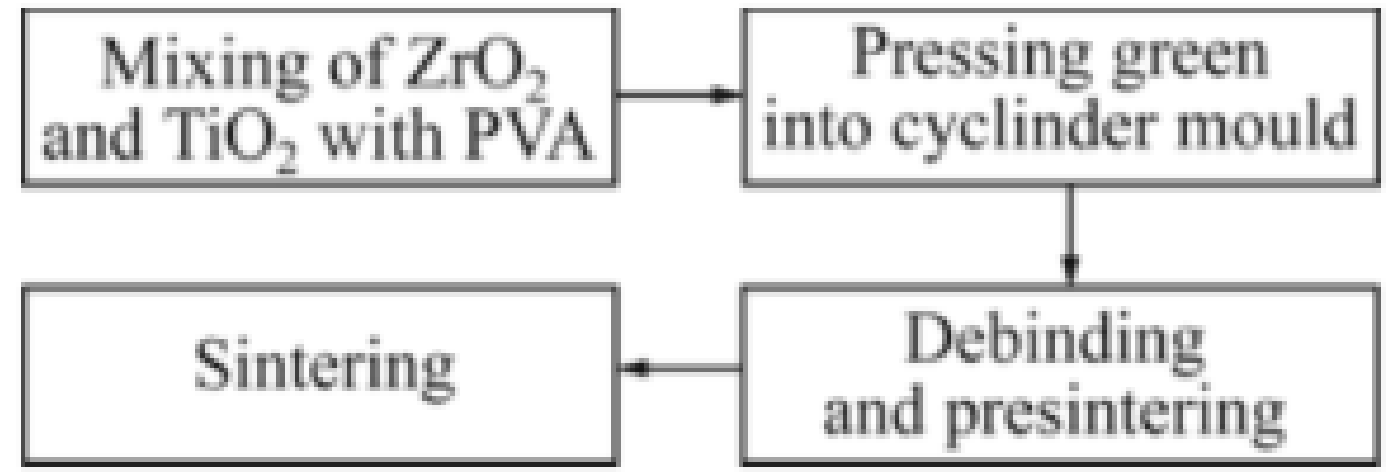

*Figure 7: Process schematic of sintering $ZrO_2$ – $TiO_2$ cylinder. Credits: [25]*

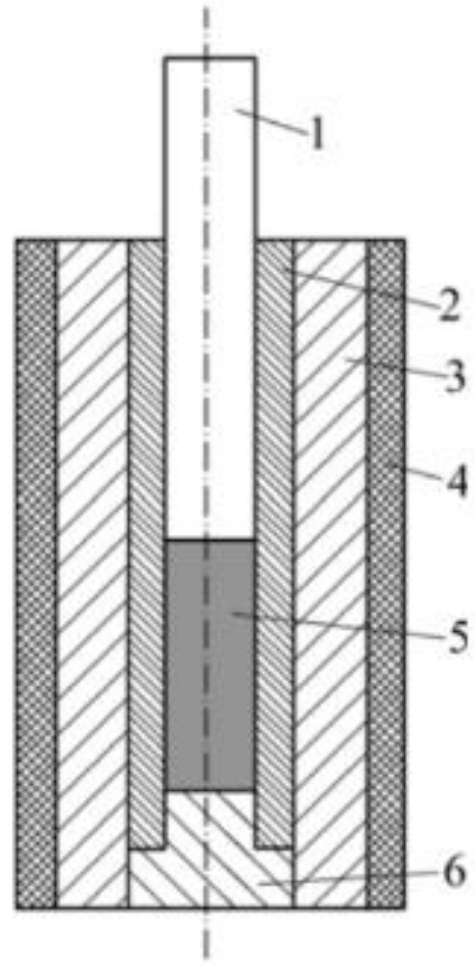

*Figure 8: Schematic structure of SPF die – 1 Formed punch; 2 Female die; 3 Outer sleeve; 4 Heating apparatus; 5 Power lot; 6 Filling piece. Credits: [26]*

To find the optimum constituent fraction for the die to match the CTE of Ti-6Al-4V at the superplastic temperatures, volume fraction of $TiO_2$ was varied and corresponding CTEs were measured. Furthermore, the influence of relative density on CTE was also studied and it was observed that higher the relative density, more is the CTE at the superplastic temperature which can be seen below in *Figure 9 (a)* and *9 (b)* respectively. Therefore, by trial and error, the optimal mixed powder with properties of 27% $TiO_2$ volume fraction sintered at 1500°C for 2 hours with a linear CTE of $8.92\times10^{-6}$ $C^{-1}$ was selected as it matched the CTE of Titanium at 900°C. This can be seen in *Figure 10* which shows the variation of volume fraction and relative density, together, based on different linear CTE to match the linear CTE of Ti-6Al-4V, which is 8.92 x $10^{-6}$ °C at 930°C.

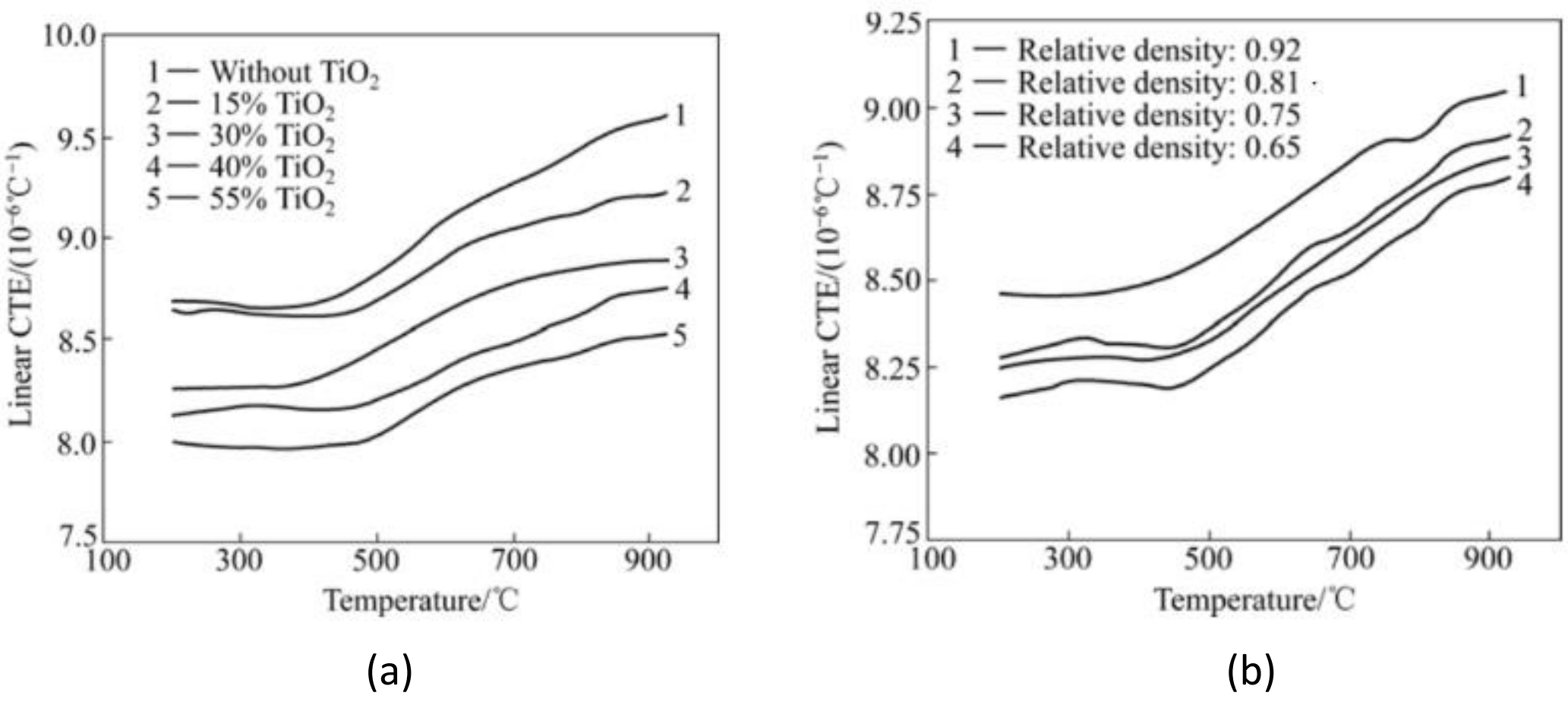

Figure 9: Relationship between linear CTE and (a) volume fraction of $TiO_2$ and (b) relative density. Credits: [25], [26]

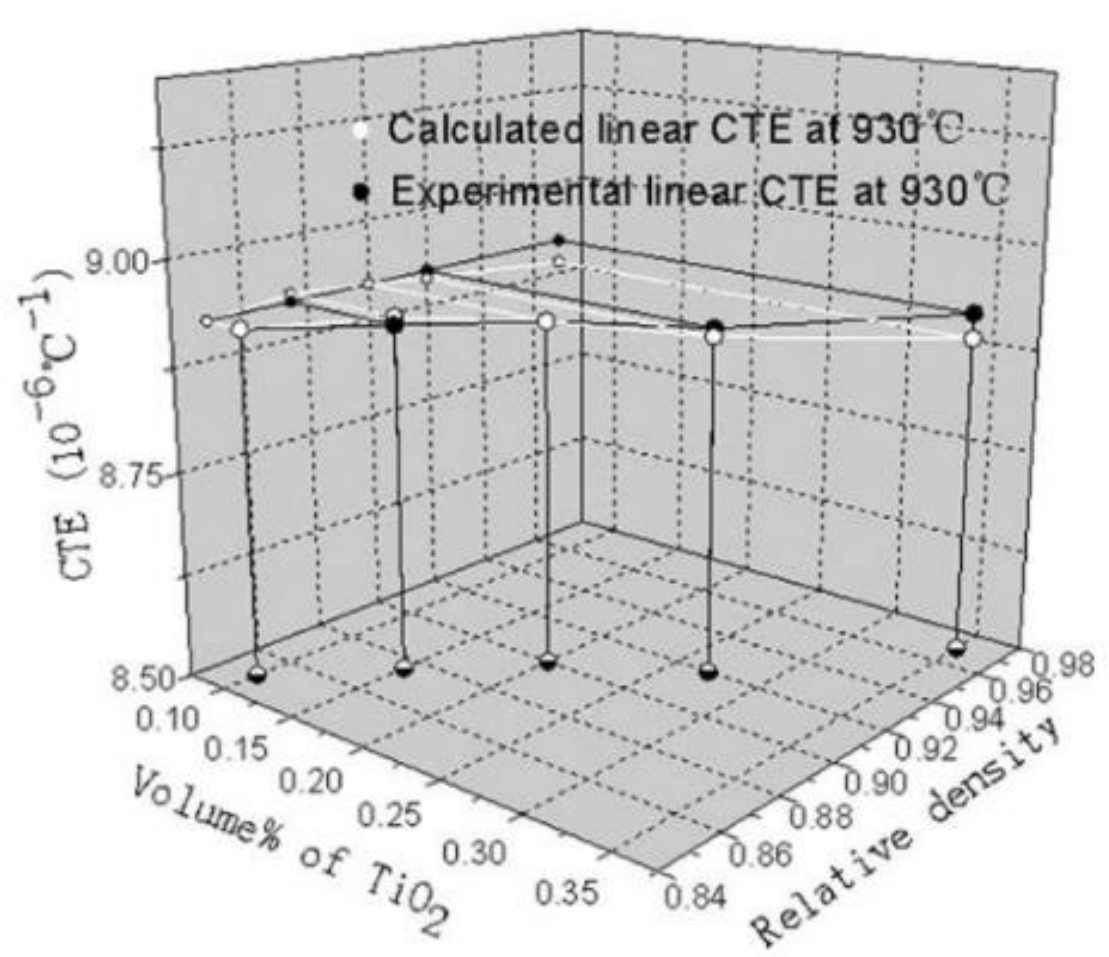

Figure 10: Different volume fraction of $TiO_2$ and corresponding relative density for manufacturing the optimal $ZiO_2$-$TiO_2$ ceramic die. Credits: [25], [26]

This mixed powder is formed into a die by pressing into a steel model at 120°C with a double cling film layer for separation in *Figure 11a*. Later, the newly formed die is sintered at 1500°C to increase strength by heating at 8°C/min and holding at maximum temperature for 2 hours before cooling to room temperature. The final die can be seen in *Figure 11b.*

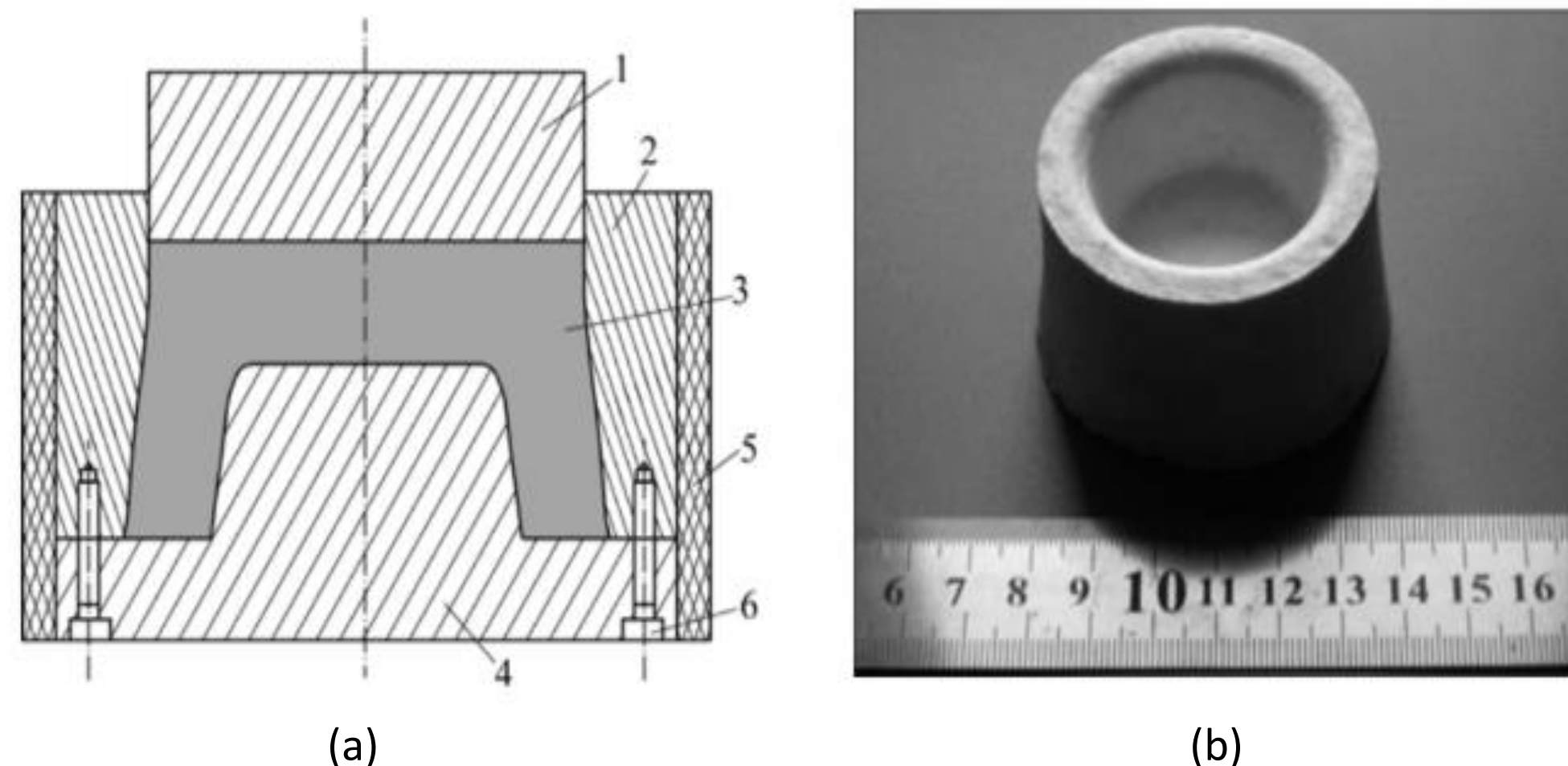

(a) (b)

*Figure 11: (a) Steel model for forming semi-finished $ZrO_2$-$TiO_2$ deep cylinder ceramic die: 1 Pressure head, 2 External mould, 3 Mixed power, 4 Internal mould, 5 Heating apparatus, 6 Screw bolts; (b) Finished $ZrO_2$-$TiO_2$ deep cylinder ceramic die. Credits: [25], [26]*

3. Tubular blank fabrication was carried out by cutting and bending of the tube [28] welding the tubular blank and checking for defects via radiography followed by sizing and welding cover with gas entrance connection. The tubular blank was bended by stress relaxation at 650°C with a holding time of 30 min followed by furnace cooling. Due to welding deformations, sizing of the pipe was to be performed at 700°C with a holding time of 45 min and was cooled similar to the tubular blank. Next, tubular blank was welded with a square butt and no gap by automatic plasma arc welding(A-PAW). During the process, the blanks were cleaned by emery paper coated with H2O/HNO3/HF solution (76/20/4 vol.%) and degreased with acetone before welding. Further details regarding welding are given in *Table 2* below.

*Table 2: Process parameters for different welding positions for manufacturing of the tubular blank. Credits: [28]*

| Welding Position | Welding Method | Arc Current(A) | Arc Voltage(V) | Travel speed (m/s) | Argon flow (L/min) |
|---|---|---|---|---|---|
| Covers | M-GTAW | 35 | 18 | 0.1 | 20 |
| Gas entrance connection | M-GTAW | 30 | 20 | - | 20 |
| Tubular blank | A-PAW | 40 | 20 | 0.06 | 25 |

### 3.1.2 Multi Step Dies

1. Two step rectangular die enclosed in a sealed die with uniform pressure applied by the inlet of pressurized inert gas such as Argon. The to-be-formed diaphragm is rigidly clamped around the periphery outside the enclosure resulting in significantly thinner border [27]. The gas pressure for blow forming is maintained between 1-2 MPa to give a dome with near-hemispherical shape along with an equilibrium biaxial stress condition at the tip and a plane-strain balanced state at the equator [23]. The variation of geometry of the workpiece in various tages of forming is shown in *Figure 12.*

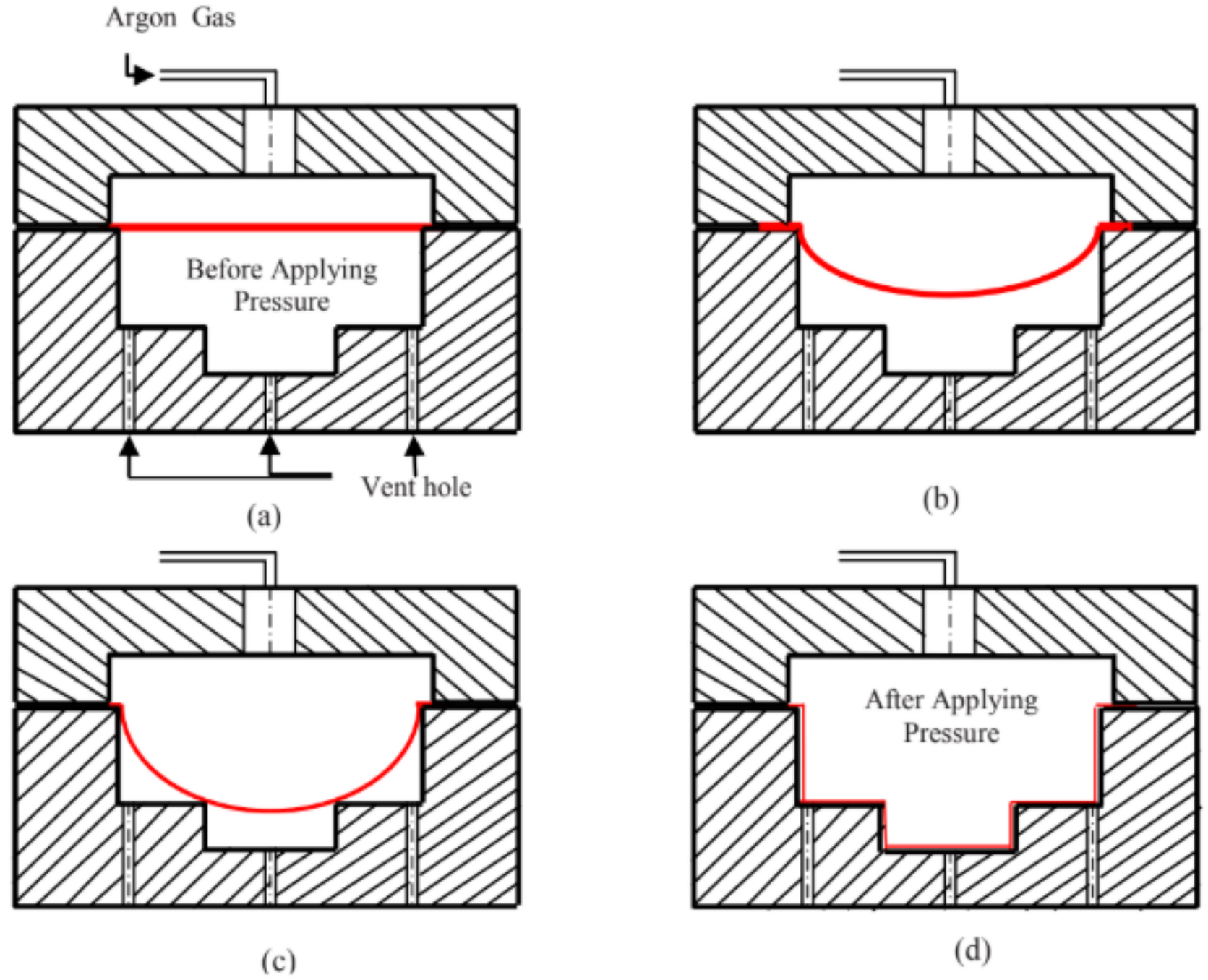

*Figure 12: (a), (b), (c) and (d) Different stages of pressure blow forming technique in a stepped rectangular die. Credits: [23]*

2. Superplastic forming is an effective process for fabricating hollow fan blades, honeycomb structures, bellows expansion joints, and large engine components [17]. However, due to slow forming rate and high cost of materials, a new hybrid sheet-forming process was used which combined deep drawing and blow forming in one step resulting in a faster forming process. The major components of the experimental setup include a three-stepped die with 80, 135 and 150% cumulative surface expansion ratios, a blank holder with INCONEL disc-springs, a 43mm stroke punch and a sealing system for blow-forming, all of which are shown below in *Figure 13.*

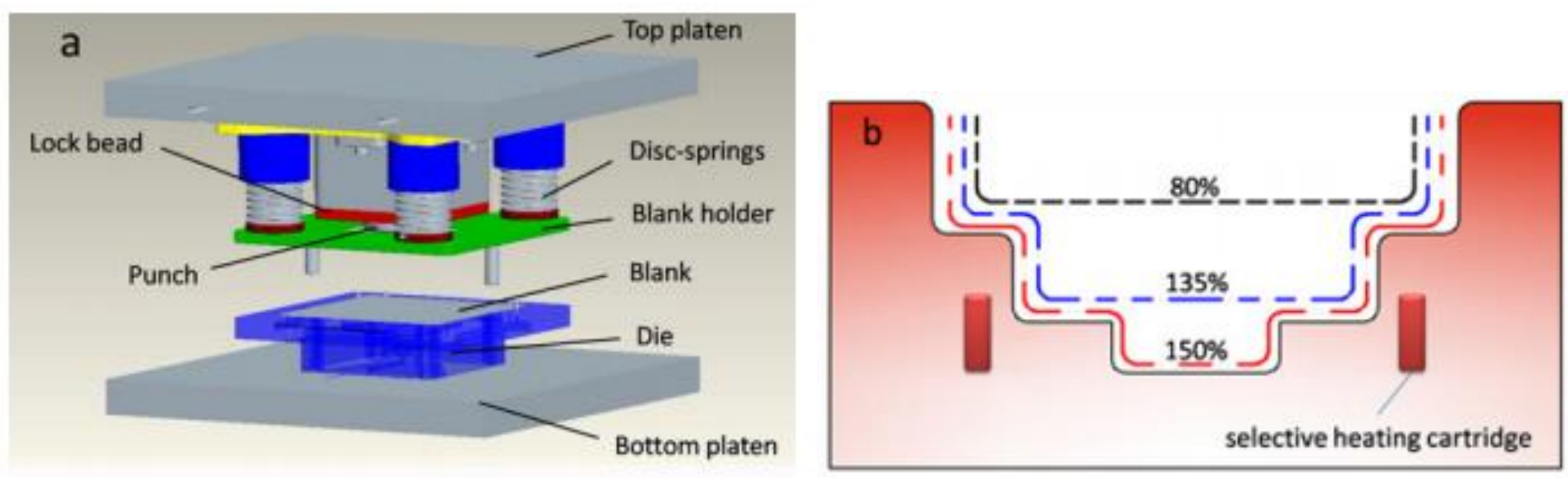

*Figure 13: (a) Components of the superplastic-like forming die assembly (b) Die-cavity with non-isothermal heating and showing % surface area expansion at each step. Credits: [17]*

### 3.1.3 Non-traditional dies

1. To make an airtight closed chamber for non-traditional die type of superplasticity, an open cylinder of Ti-6Al-4V with circular base (2.38 mm wall thickness and 16.6 mm length with 2.24mm thick disk base) was machined and sealed with another disk-like top of Ti-6Al-4V/TiC composite of 2.24mm thickness [23]. To this, a 6.35mm diameter pipe of Ti-3Al-2.5V was welded to allow gas flow and K-type thermocouple assemblies inside the sealed cylinder. For determining the gas-pressure induced radial and circumferential strains, the disks were marked with concentric circles spaced 1 mm apart and parallel lines going through the center of the disk respectively. A Boron-Nitride (BN) coated 2.5mm thick sleeve to prevent radial strains and allow circumferential strains housed the closed cylinder. The entire assembly, as in *Figure 14*, was placed inside a cylindrical graphite chamber within a stainless-steel vessel.

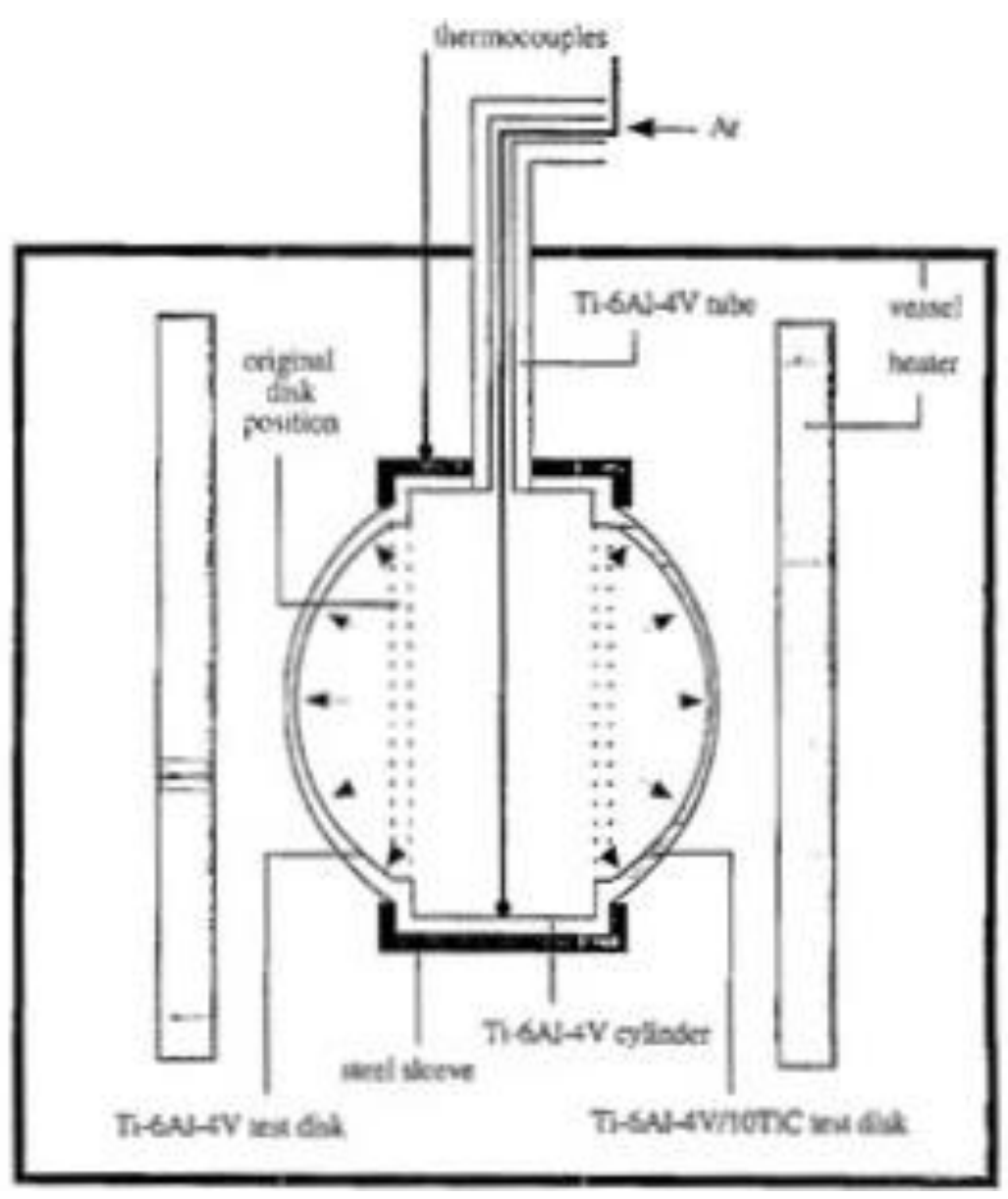

*Figure 14: Complete assembly of an airtight closed chamber for SPF forming. Credits: [23]*

2. Before forming, the sheets underwent heat treatment up to 1070℃ in an induction heating machine of 25 kW power [7]. Next, they were air-cooled and each sheet was cut according to the circular clamping periphery. The forming vessel is made up of a high temperature nickel alloy with a removable top (A). The top houses a stainless-steel die (B) placed inside the lower vessel (C). Heat is supplied through the resistance-heating coils (D) and constant superplastic temperature of 920℃ is maintained by a PID (proportional-integral-derivative) controller while Argon gas pressure is regulated by computerized mass-flow valves. The entire assembly setup is given in *Figure 15* below.

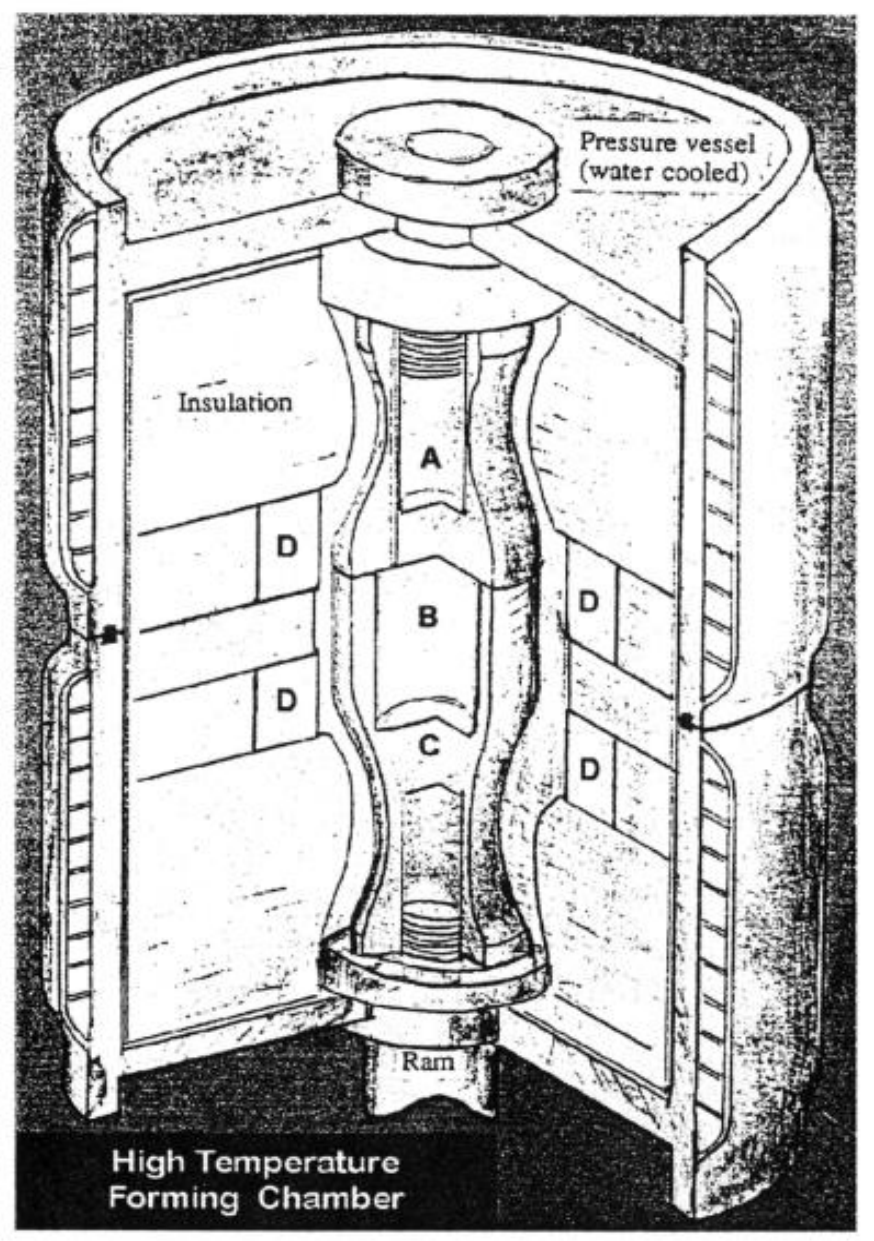

*Figure 15: Schematic outline of the high temperature forming chamber. (A) – Top pressure vessel (B) Die housing (C) Bottom pressure vessel (D) Heater. Credits: [7]*

## 3.2 The SPF Process

### 3.2.1 With Ti-6Al-4V

1. The structure of setup was divided into pre-forming die, sheet metal and forming die [28] as shown in *Figure 16 (a)*. Two vent holes at the top (1 and 2) and one at the bottom (3) were provided to prevent the sheet sticking to the forming die. The Ti−6Al−4V sheet was heated at a rate of 15 ℃/min up to 930 ℃ and held for 60 minutes before pre-forming. Then, the high-pressure nitrogen gas ramped up to 1.4MPa was used to blow the sheet into the pre-forming die in 750 seconds. For achieving adaptation to the deformed state, a 30-minute holding time was observed following which more nitrogen was used to blow the sheet into the forming die at 1.4MPa in 1500 seconds after which a dwelling time of 1500 seconds was provided. The entire variation of pressure vs time for the process is shown in *Figure 16 (b).* This was achieved by a two-step blowing process: first the gas was blown through vent hole 3 at pre-forming pressure and then it was released through vent hole 1 and 2 while gas at forming pressure was injected through vent hole.

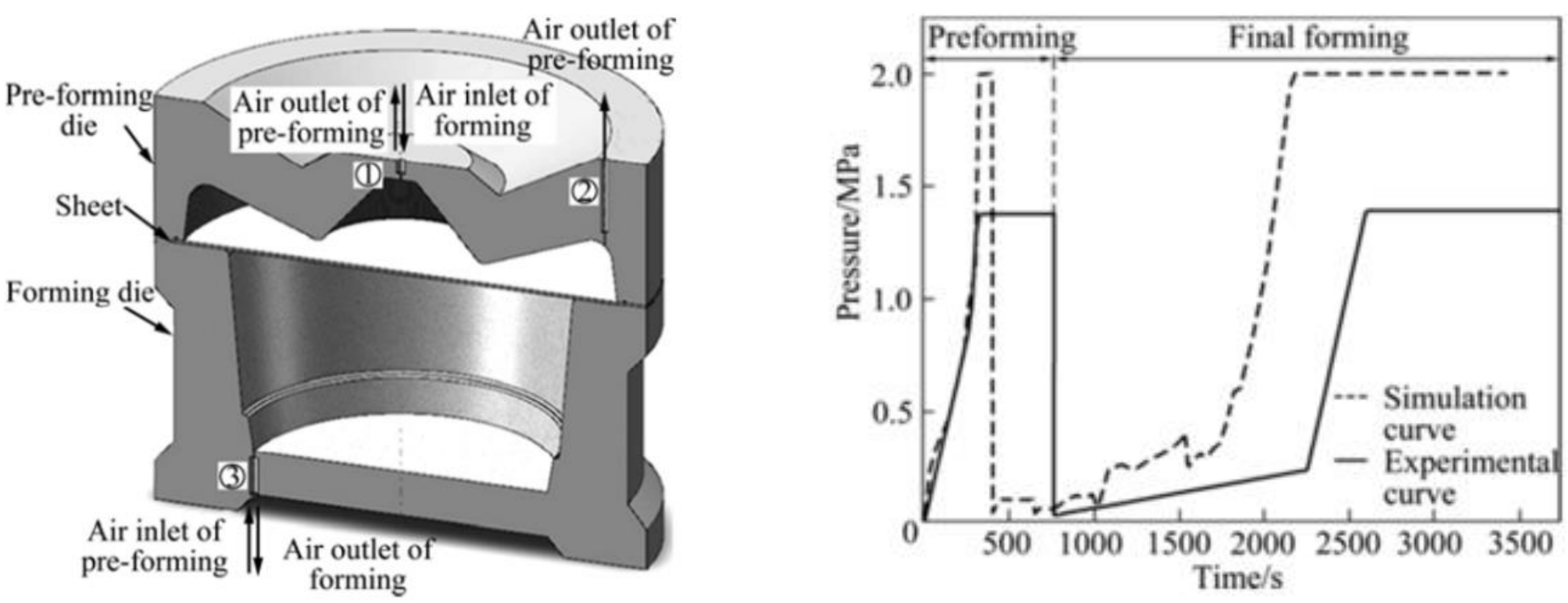

*Figure 16: (a) Schematic structure of the die for SPF (b) Experimental Pressure vs Time curves. Credits: [28]*

2. For the actual SPF process shown in *Figure 17 (a), (b), (c) and (d)*, the tubular blank, coated with anti-oxidizing agent, and die were pre-heated to 150-200℃ while being lubricated by graphite powder at their contact surfaces [24]. The entire apparatus was kept in a furnace at 927℃ (the approximate superplastic temperature for Ti-6Al-4V) for 30 minutes. Initially, gas at 0.2MPa was injected into the tubular blank for 30 minutes to initiate circumferential plastic deformation followed by lowering of the punch to close the upper, middle and lower dies tightly. Subsequently, gas at 2.5MPa was injected for a period of 10 minutes after which the punch was removed and the setup was kept in room temperature to cool. Following sufficient cooling, the middle die was removed and the two covers were ground from the formed alloy. The bellows were then machined to the required dimension by lathe and all the applied coatings were removed by grit blasting. A digital micrometer was used to measure the thickness at various points along the finished bellows and it showed variation of about 9% which can be seen in *Figure 17 (e)* below.

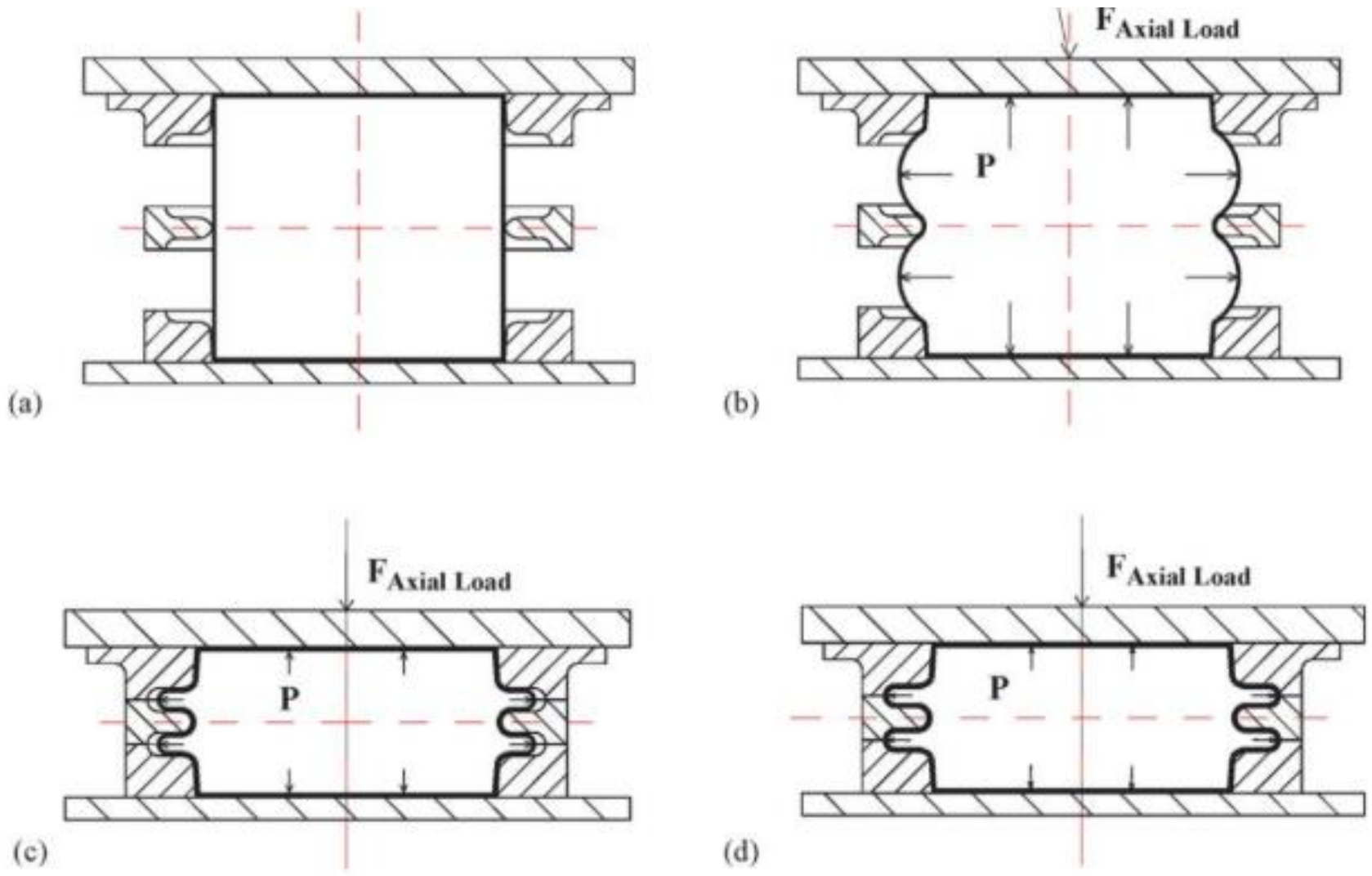

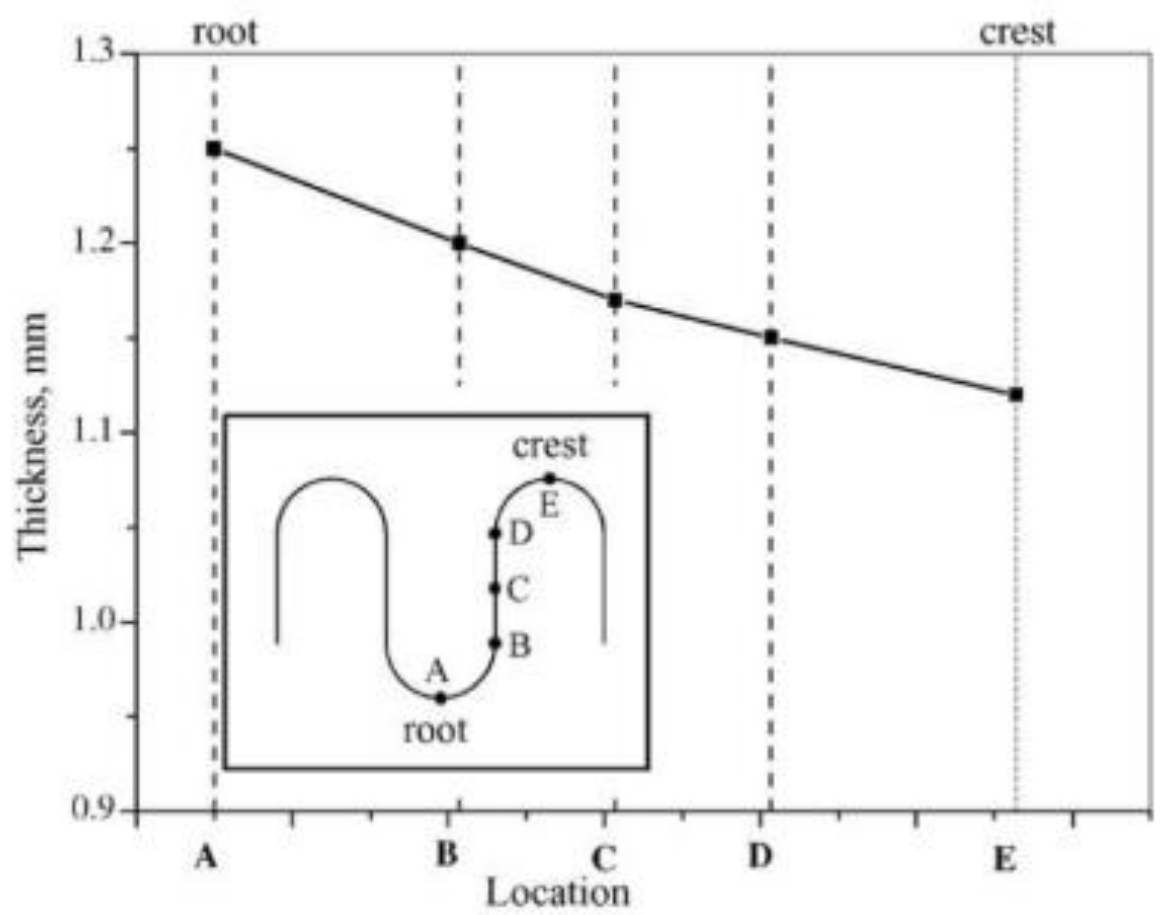

*Figure 17: SPF procedure: (a) original state (b), bulging (c) dies clamping, (d) shape forming, (e) Thickness distribution of as-formed bellows. Credits: [24]*

3. The honeycomb structure [29] is produced by selective bonding of four titanium sheets, each with a specific core pattern, and placing them one over the other. Core flat sheets have a thickness of 0.3mm and face flat sheets have a thickness of 0.8 mm. To ensure that no gas leaks out, all the four sheets are sealed along their periphery and placed into process tooling where they are heated to the SPF temperature. Moreover, to reduce post-processing, close tolerances and tight quality control measures should be implemented during the entire process. The four sheets are subjected to forming by applying gas pressure until they come in contact with the die [29]. Pressurized gas is injected until diffusion bonding of the sheets resulting in one, monolithic structure is completed. Optimum amount of pressure needs to be applied to ensure superplastic forming at stop-off areas. Its variation with time as shown in *Figure 18 (a)*.

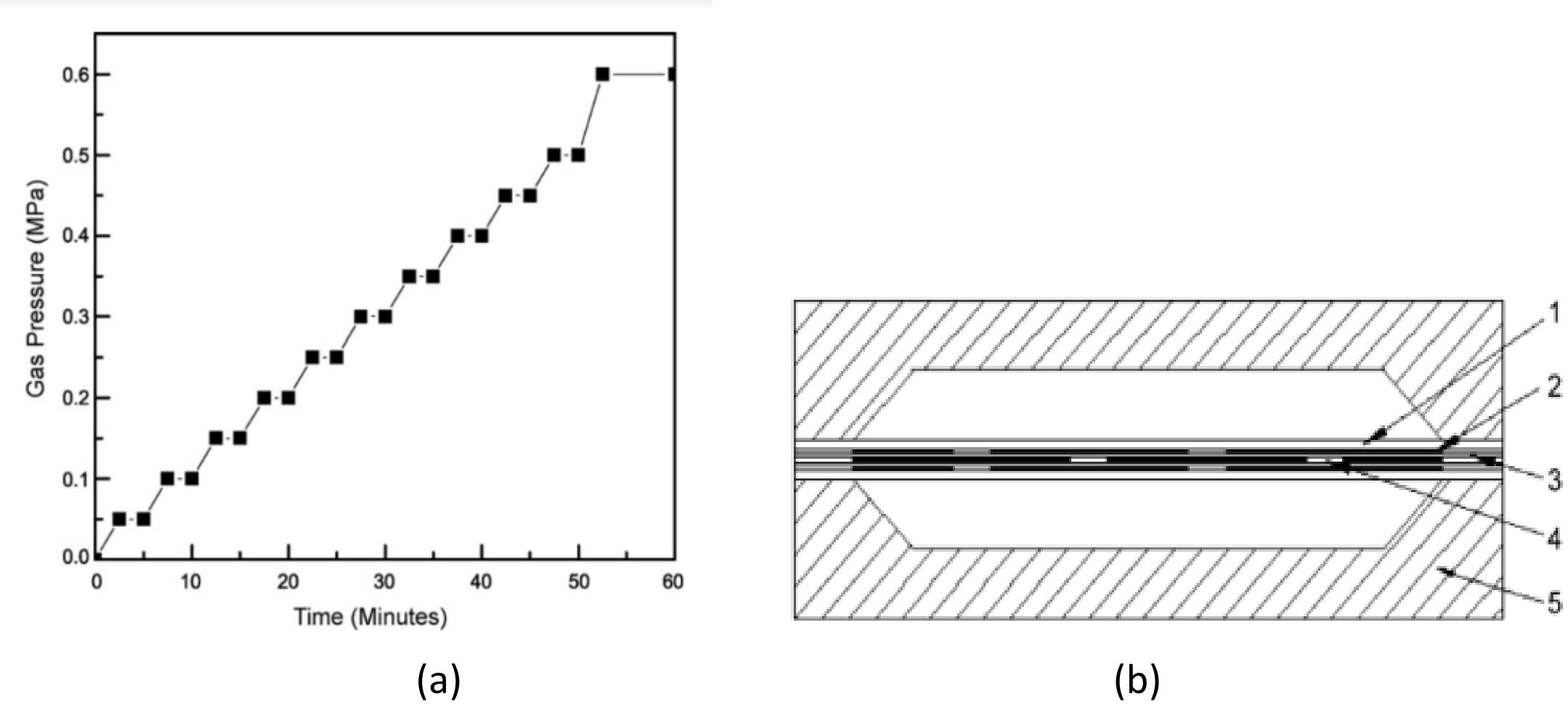

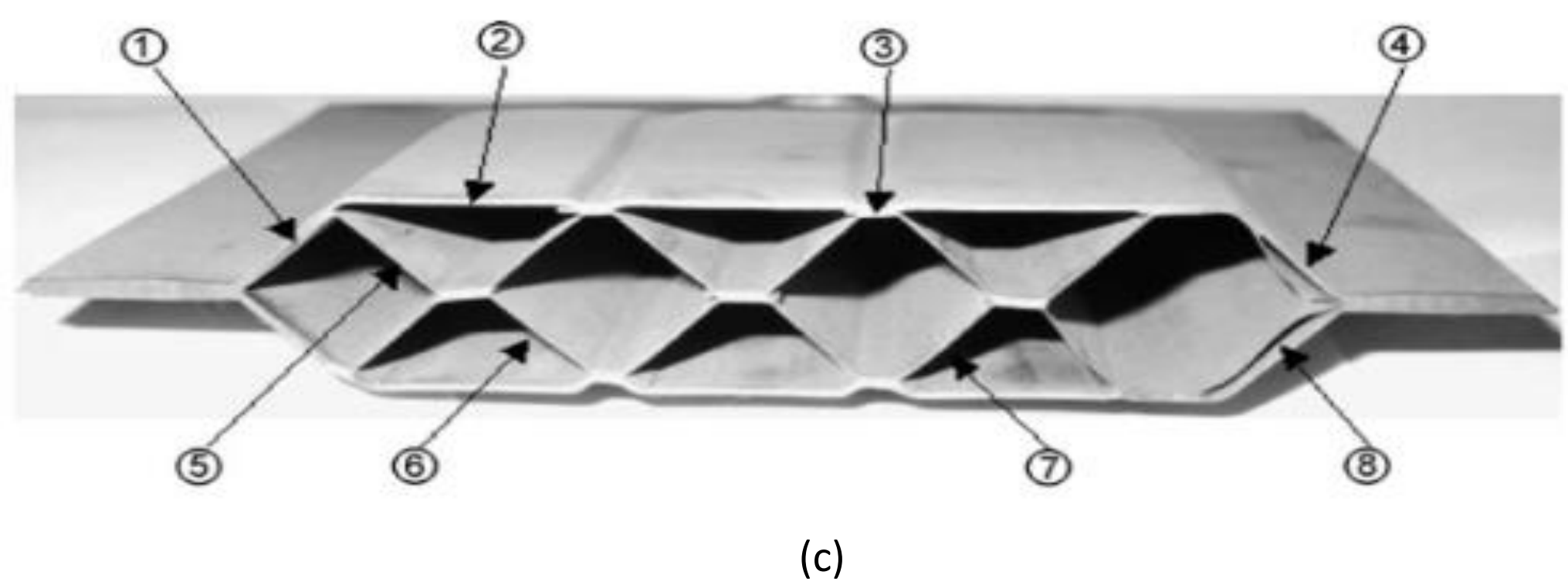

(c)

*Figure 18: (a) Pressure vs time variation for SPF-DB processing (b), Assembly for the processing. 1 – face sheet, 2 – stop off, 3 – core sheet, 4 – stiffening rib, 5 - die (c) Points at which thickness was measured whose values are shown in Table 3. Credits: [29]*

*Table 3: Thickness distribution of the honeycomb structure. Credits: [29]*

| Location | 1 | 2 | 3 | 4 | 5 | 6 | 7 | 8 |
|---|---|---|---|---|---|---|---|---|
| Thickness (mm) | 0.76 | 0.77 | 0.78 | 0.76 | 0.28 | 0.28 | 0.28 | 0.28 |

4. Two categories of sheets: as received and heat treated were subjected to forming at the above conditions for 1000 seconds with strain rates ranging from $1{\times}10^{-4}$ to $8{\times}10^{-4}$ $s^{-1}$ [7]. After cooling in atmosphere, thickness was measured for both types of sheets. The lower cup showed large reduction in thickness and greatest thinning was found at the innermost corner for both the categories. However, in heat-treated sheets, there was significantly higher thinning in the non-transformed regions than the transformed regions due to the interface effect between the two regions. Comparison with FE simulations is given below in *Figures 19 (a) and (b)* in which length is the peripheral distance along the formed shape direction out from the center.

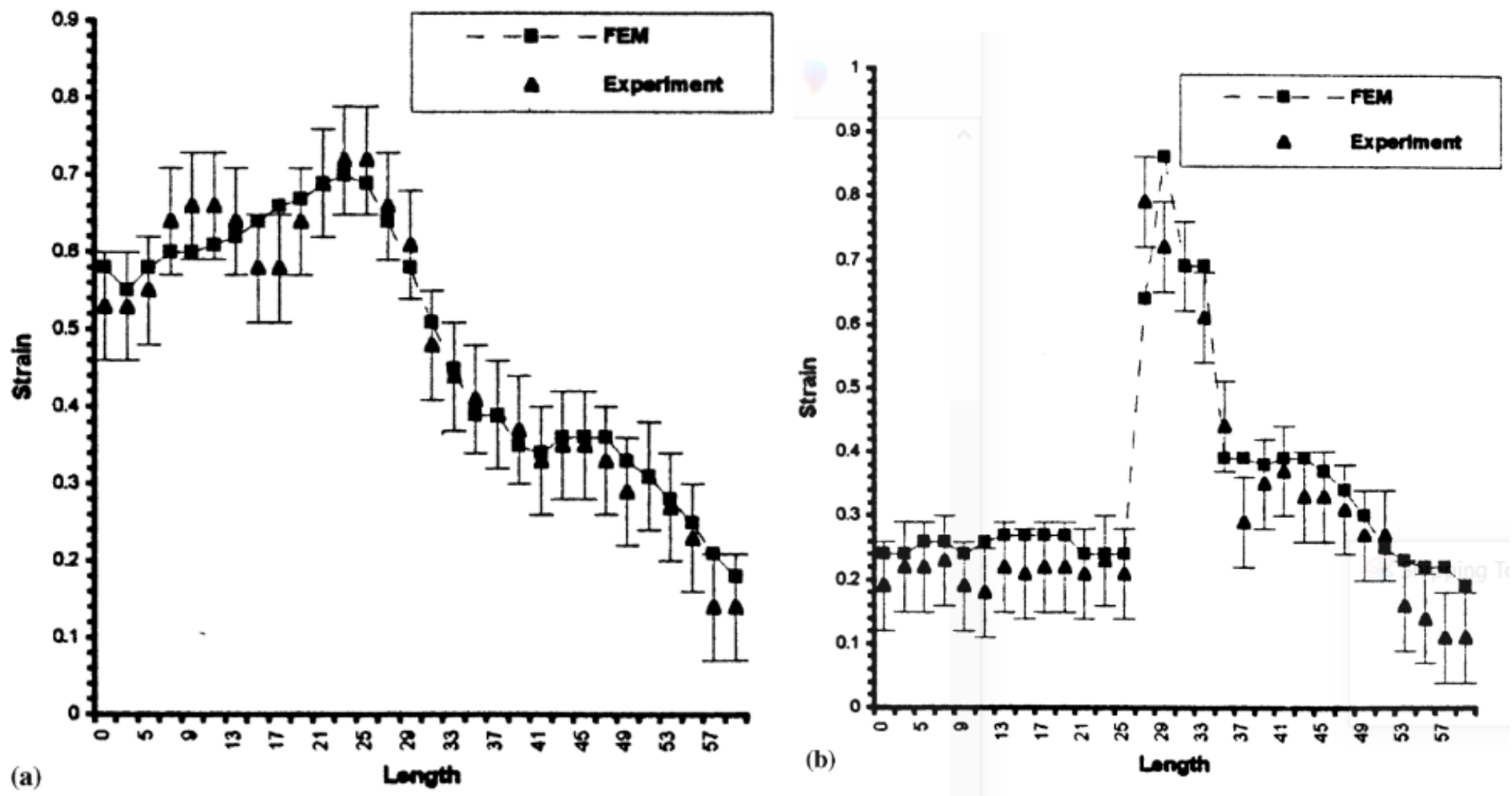

*Figure 19 Comparison of finite element simulation and experimental strains for (a) the as-received blank and (b) the blank with transformed structure. Credits: [7]*

5. To carry out SPF, the chamber and vessel initially were maintained at 10 mTorr and evacuated again to the initial pressure conditions [23]. To achieve degassing at 10 mTorr pressure, the chamber was heated to 150℃ for 10 h and ramped to 350, 600 and 800℃ with a holding time of 1 h at each increment. For the isothermal experiment at 1000℃, 99.9996% pure Argon gas at 350 kPa and 120 kPa was introduced in the test chamber and vessel respectively. Titanium foil getters were used to prevent oxidation of the sample and the experiment was conducted till failure of the Ti-6Al-4V weld. For the cyclic experiment, Argon gas at the same respective pressures was used but with temperature range of 580℃ and 1020℃ with a holding period of 600 s (with ramps of about 60 s) at each increment. These cyclic experiments were conducted till the Ti-6Al-4V dome apex developed four cracks of approximately 1 mm length. Constant pressure was maintained via an Argon reservoir and two sets of cyclic experiments were conducted. In the first, both the reinforced and unreinforced disks were allowed to deform with interruptions after every 10 hours for visual inspection while in the second, unreinforced disk was supported by a steel insert to prevent deformation, allowing only the reinforced disk to deform. Point micrometer was used to calculate the principal strains by measuring the thickness of the formed dome at different distances from the apex.

Superplastic forming is not only carried out for Ti-6Al-4V, but there are many other titanium alloys which are used too. These other alloys usually have small proportions of vanadium or molybdenum to impart a few specific desired properties to the alloy and the formed component. The nuances of the forming of these other titanium alloy has been explained in the next subsection.

### 3.2.2 With Other Titanium Alloys

1. Welded envelopes of VT6S (Ti–6Al–4V), VT14 (Ti–4.2Al–2.7Mo–1.2V) and OT4-1 (Ti–1.8Al–1.45Mo) of thicknesses 0.8, 1 and 5 mm respectively were used for superplastic free forming of a spherical shell [16]. For manufacturing the spherical envelope, two circular sheets were taken and a pipeline for Argon entry was allowed for by welding a gas entrance connection to one of the sheets in the middle. Their properties are tabulated in *Table 3* along with the results in *Table 4,* in which $D_0$ refers to original sheet diameter while $D_s$ signifies the diameter of the formed sphere. Six sheets were welded in pairs across their perimeter based on the relative rolling direction - 0°, 45° and 90°. Three methods namely argon arc welding, roll seam welding and diffusion bonding were used for the process. To allow for SPF, the fastened envelope was placed on a π-type stand in a furnace. Argon was injected into the envelope through the provided pipeline. Gas pressure was controlled according to the predetermined pressure–time cycle for optimum conditions for SPF. During forming, geometry of the envelope was monitored visually resulting in an error of 3 mm and via a graduated scale on a computer fitted with a camera resulting in an error of 1 mm. *Figure 20* shows the final spherical shell formed.

*Table 4: Material properties of the Titanium sheets used in manufacturing spherical shells out of welded Titanium envelopes. Credits: [16]*

| Chemical Composition | Initial thickness $s_0$ (mm) | Average grain size d* (μm) | Optimal conditions for superplasticity | | | Material constants | | Procedure used |
|---|---|---|---|---|---|---|---|---|
| | | | $T_{opt}$ (℃) | $S_{opt}$ (MPa) | $X_{opt}$ ($s^{-1}$) | K (Mpa $s^{-m}$) | m | |
| Ti-6Al-4V | 5 | 3-5 | 920 | 15 | $2 \times 10^{-4}$ | 450 | 0.40 | Standard |
| Ti-4.2Al-2.7Mo-1.2V | 0.8 | 1-2 | 870 | 5 | $3 \times 10^{-3}$ | 100 | 0.52 | Standard |
| Ti-1.8Al-1.45Mo | 1 | 3-5 | 890 | 20 | $5 \times 10^{-4}$ | 290 | 0.35 | Standard |

*Table 5: Results of experiments on superplastic free forming of spherical shells out of welded titanium envelopes. Credits: [16]*

| Sheet material | Chemical composition | Thickness (mm) | $D_0$ (mm) | $D_s$ (mm) | $D_0/D_s$ | Deviation (mm) |
|---|---|---|---|---|---|---|
| VT6S | Ti-6Al-4V | 5 | 200 | 164 | 1022 | ±2 |
| VT6S | Ti-6Al-4V | 1 | 108 | 89 | 1.21 | ±2 |
| VT6S | Ti-6Al-4V | 1 | 188 | 158 | 1.19 | ±2 |
| VT6S | Ti-6Al-4V | 1 | 290 | 246 | 1.18 | ±3 |
| VT6S | Ti-4.2Al-2.7Mo-1.2V | 0.8 | 188 | 158 | 1.19 | ±2 |
| OT4-1 | Ti-1.8Al-1.45Mo | 1 | 188 | 160 | 1.18 | ±2 |

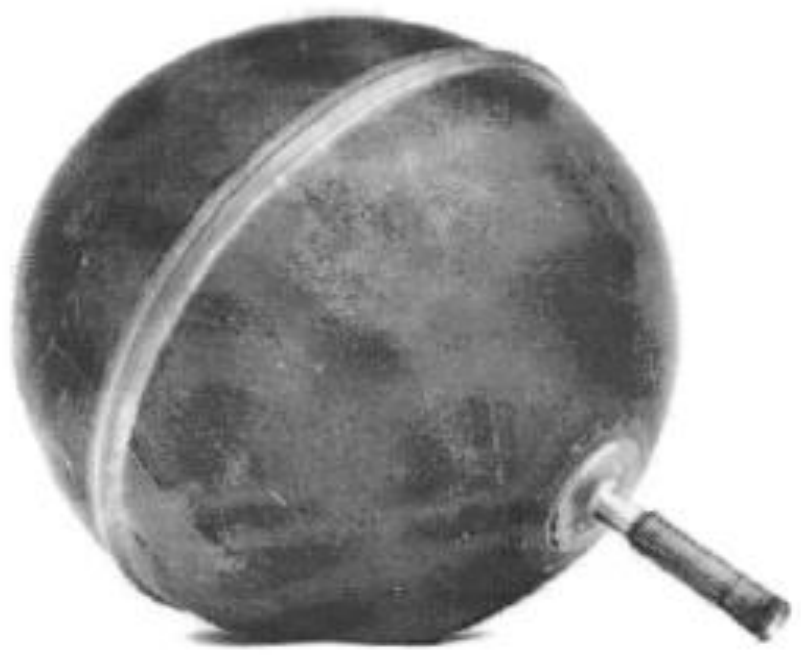

*Figure 20: Spherical shell formed out of the envelope manufactured by diffusion bonding ($D_0$ = 200 mm, $D_s$ = 164 mm). Credits: [16]*

## 3.3 Process Parameters

1. In addition, it was found that splitting the die into a pre-forming and forming die as in *Figure 21* and separately modifying their surfaces resulted in achieving greater uniformity of sheet thickness as compared to the conventional single-piece die blow forming [29]. With this regard, increasing friction of pre-forming die by mechanical machining and decreasing the friction of forming die by spraying BN powder to obtain more accurate finished products. The friction coefficient was monitored by compression ring method and its values for machined surface and BN powder sprayed surface were obtained as 0.55 and 0.15 respectively.

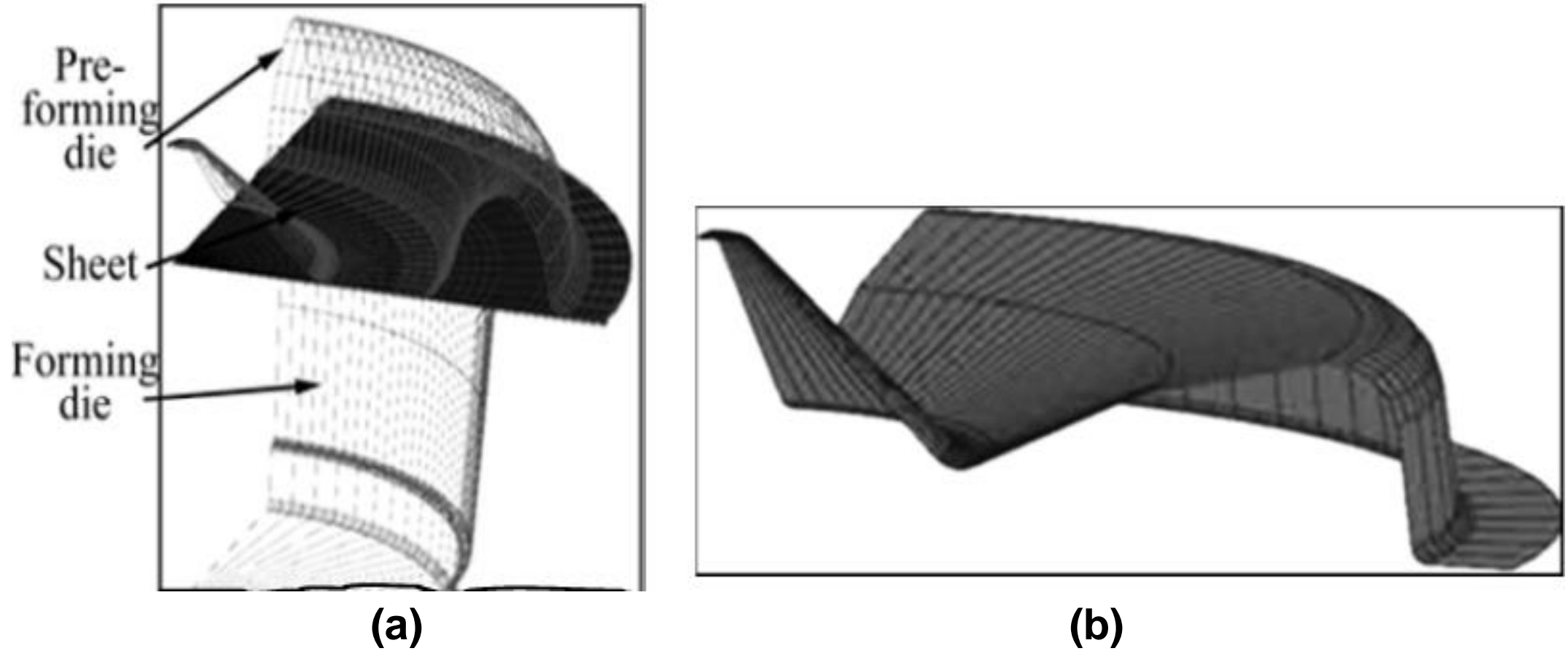

*Figure 21: (a) Splitting of die into forming and preforming results in uniform thickness distribution (b) Meshed geometry showing the shape of preforming die. Credits: [29]*

2. Uniaxial tensile texts were conducted on Ti–22Al–24Nb [30], with traces of Mo, V and Si on Instron 5500R, an electronic universal material testing machine. Since both temperature and strain rate affects the superplastic ability, first, optimal temperature was set by conducting tensile tests within 900-980°C at increments of 30°C at a strain rate of .0001 $s^{-1}$. Next, optimal strain was selected at .0005 $s^{-1}$ for proceeding study. The superplastic forming tests were conducted in a 1000kN forming machine via gas bulge forming until the sheet fitted the mould completely.

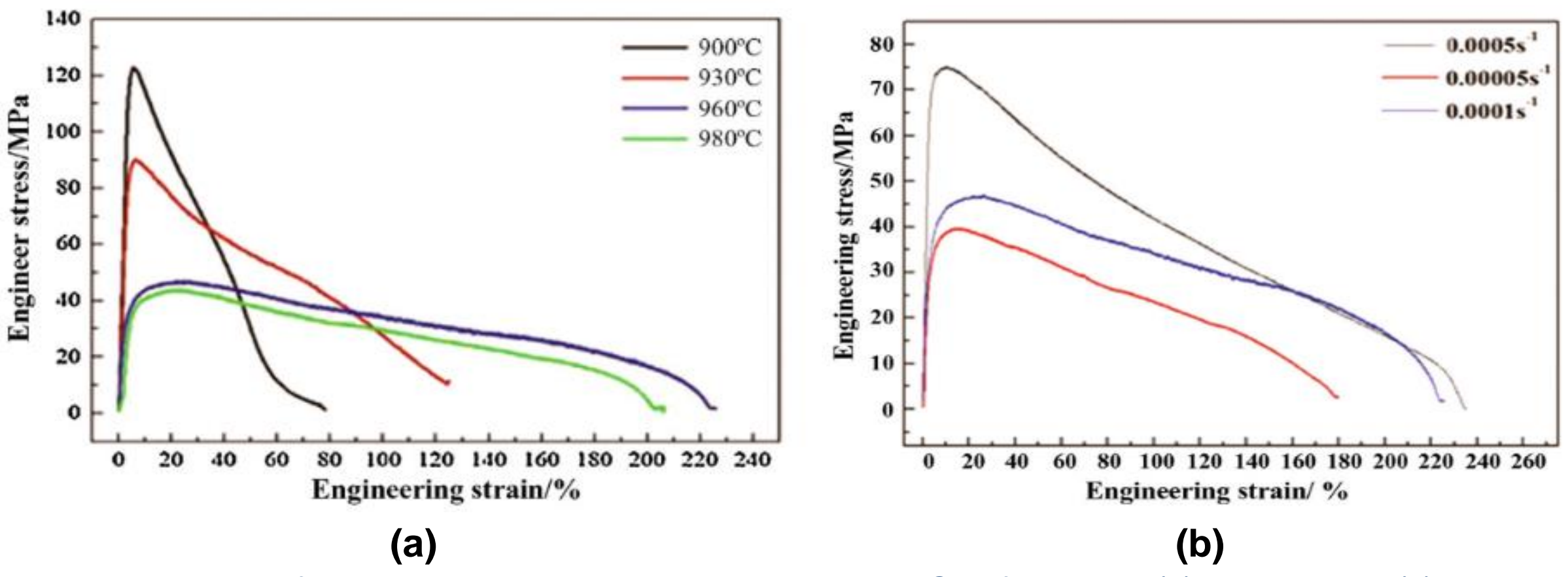

*Figure 22: Variation of engineering stress and engineering strain in SPF for various (a) temperatures (b) strain rates. Credits: [30]*

As can be seen from *Figure 22*, tensile stress increases to a rapid peak with increase in tensile strain and decays somewhat gradually in the case of 960° and 980°C. With the increase in temperature, superplasticity increases in general while material's yield strength decreases due to decrease in flow stress of the material allowing for greater deformation. As we increase the temperature beyond 960℃, oxidation effects become dominant and lead to a lower fracture strain at 980°. It is also observed that at 960℃, a strain rate of .0005s-1 allows the largest deformation and hence Ti–22Al–24Nb alloy has the optimal superplasticity at the temperature of 960 ℃ and strain rate of 0.0005 $s^{-1}$.

3. As seen above in the various procedures, clamping of the Titanium alloy sheets along their periphery during blow forming results in thinning of the clamped edges [16]. $D_0$ is the original diameter of the circular sheets while $D_s$ is the diameter of the manufactured sphere. It was found out that shells which were formed with sheets having relative rolling directions as 45° and 90° had curved welds i.e. the welds did not lie in the equatorial plane. This resulted in non-uniform plastic strains due to different length of meridians causing large deviations from the desired spherical shape. It was then concluded that symmetrical arrangement of sheets (0° relative rolling direction) was required for desired results. As seen in *Table 5*, for VT6S of 1 mm thickness, the $D_0/D_s$ ratio is 1.21 implying that the sheets 'possess' approximately 21% of plasticity resource to account for this reduction in diameter.

During the SPF, the equatorial ring containing the welded joint might lose its circular shape. Folds begin to appear in this equatorial zone during the early stages of the process but if the initial sheet thickness is small, these folds may disappear towards the end. If thick sheets or low plasticity welds are used, wrinkles continue to exist in the final formed shell. Thus, elimination of wrinkles and folds is of practical interest where high finish of the product is required.

4. Diffusion bonding combined with superplastic blow forming is the most widely used process for Ti-6Al-4V SPF as it leads to the formation of a honeycomb structure which is otherwise not producible [29]. This leads to significant reduction in cost and weight (around 30-50%) of the formed part and finds application in complex sandwich-like structures with multiple layers of the alloy.

To find the optimal conditions for diffusion bonding, variation of shear strength of the formed joint with bonding temperature, pressure and time is studied. From *Figure 24(a)* it is evident that there is in fact an optimal temperature range (1190-1210 K) for which the shear strength is maximum. This can also be observed in the SEM images taken at 1123 K and 1203 K. It is seen that micro-plastic deformation at material interface is insufficient in the former case while in the latter, due to increased surface area in contact leading to better elemental diffusion, the shear strength increases by almost 1.5 times.

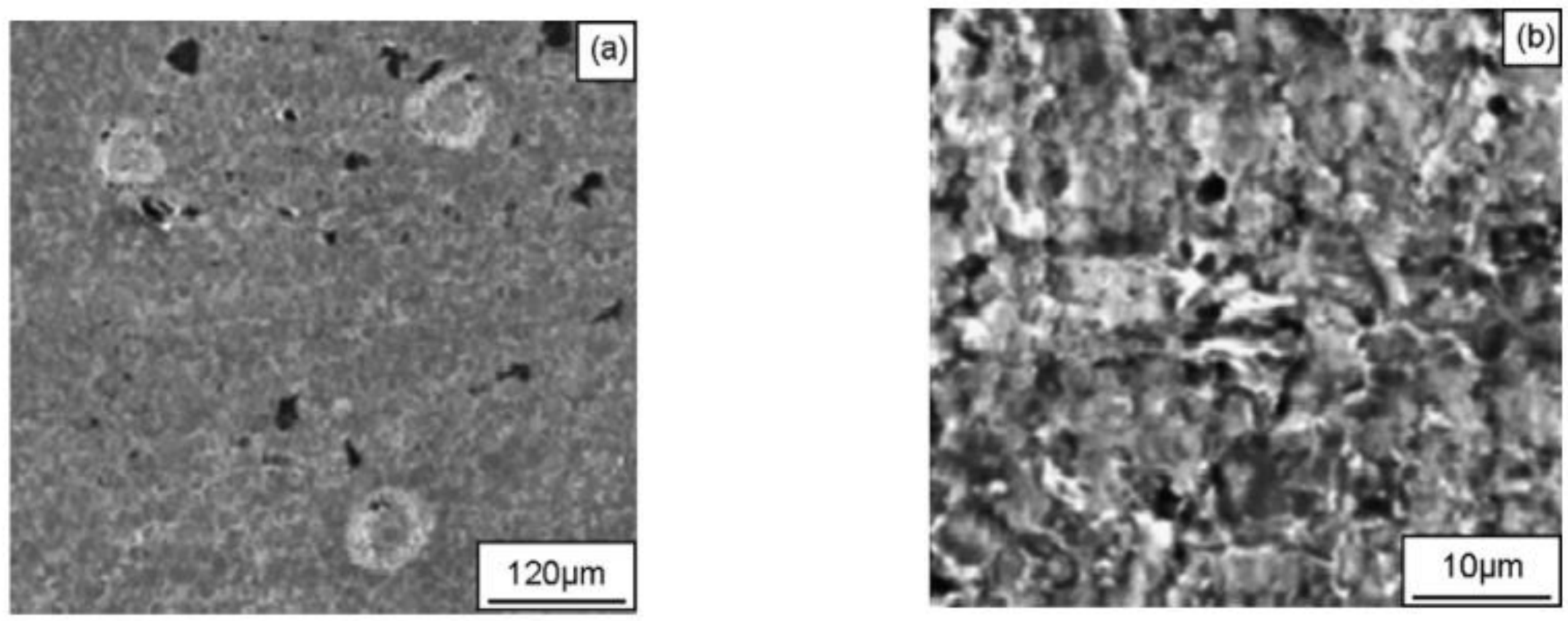

*Figure 23: SEM images taken at (a) 1123 K and (b) 1203 K showing that micro-plastic deformation is insignificant at lower temperatures. Credits: [29]*

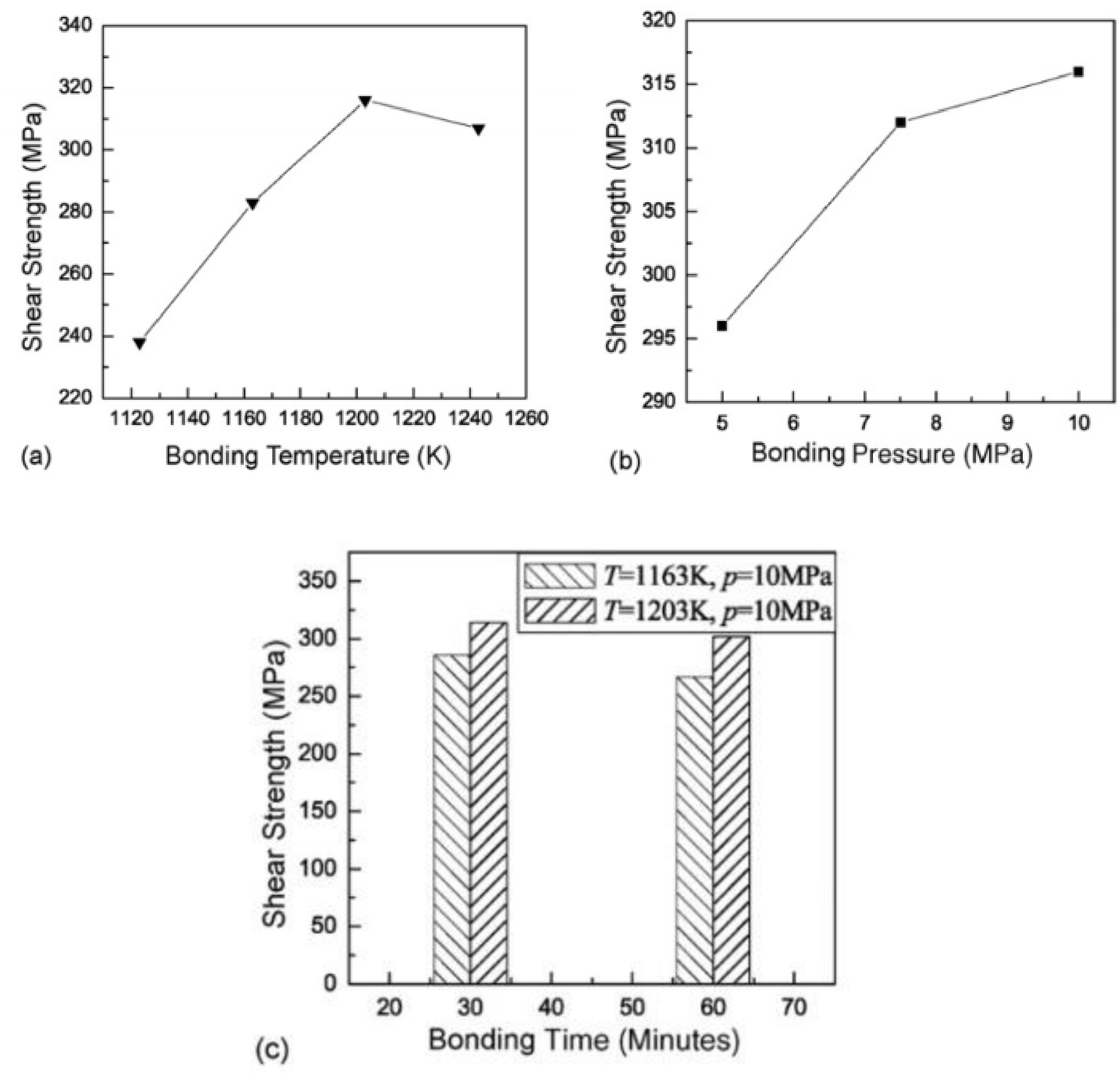

*Figure 24: Variation of shear strength and bonding (a) temperature (b) pressure (c) time during diffusion bonding combined with superplastic blow forming. Credits: [29]*

Further, as higher pressure leads to increase in contact surface area, we expect an increase in the shear strength as in *Figure 24(b)*. This phenomenon can also be seen via the graph plotted over 5-10 MPa range. Doubling the pressure from 5 MPa to 10 MPa leads to an increase of 20 MPa shear strength. Hence, effect of pressure on diffusion bonding as compared to the effect of temperature is not as large as observed in *Figure 25*. It should be noted that, indefinitely increasing the pressure will lead to large plastic deformations causing loss of precision in the structure.

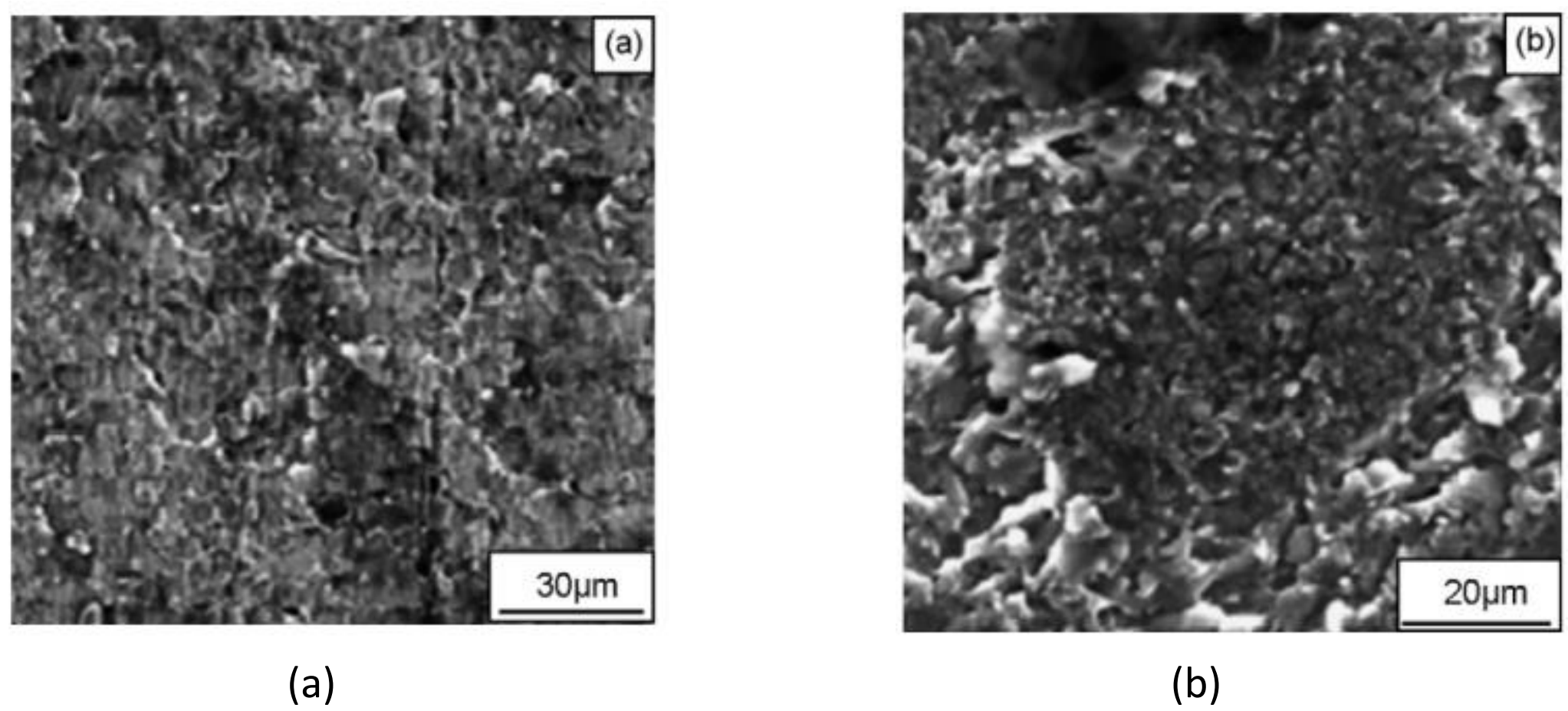

(a) (b)

*Figure 25: Microstructure at (a) 7.5 MPa and (b) 10 MPa at constant temperature showing that the effect of pressure is incomparable to the effect of temperature. Credits: [29]*

Holding time has a secondary effect on the shear strength of the joint since it cannot alter the surface area in contact. Therefore, increasing holding time at lower temperatures (1163 K) does not increase the shear strength of the joint despite increasing the elemental diffusion of the alloy. However, at higher temperatures (1203 K), due to increase in surface area in contact, increase in holding time increases the shear strength of the joint. This phenomenon resulted in an optimum time of 30 min at 1203 K and 10 MPa pressure as observed in *Figure 24(c)*

5. For both the isothermal and cycling experiments, unreinforced dome underwent a larger deformation than the reinforced dome [23]. Density measurements were performed by the Archimedes method and it was concluded that there was no significant change in these values. For similar total deformation times, it was found that thermal cycling resulted in significantly larger deformations than isothermal holding for both the unreinforced alloy and the Ti-6Al-4V/TiC composite in *Figure 26 (a)* and *(b)* respectively. It was also observed that total strain to dome fracture and the deformation rate are more in the unreinforced alloy than in the composite in spite of the larger number of cycles in the latter. The three principal strains were plotted as function of original distance from the disk center for both the set of specimens in *Figure 27*. These strain values differ significantly for each set separately due to gravity leading to an uneven dome shape.

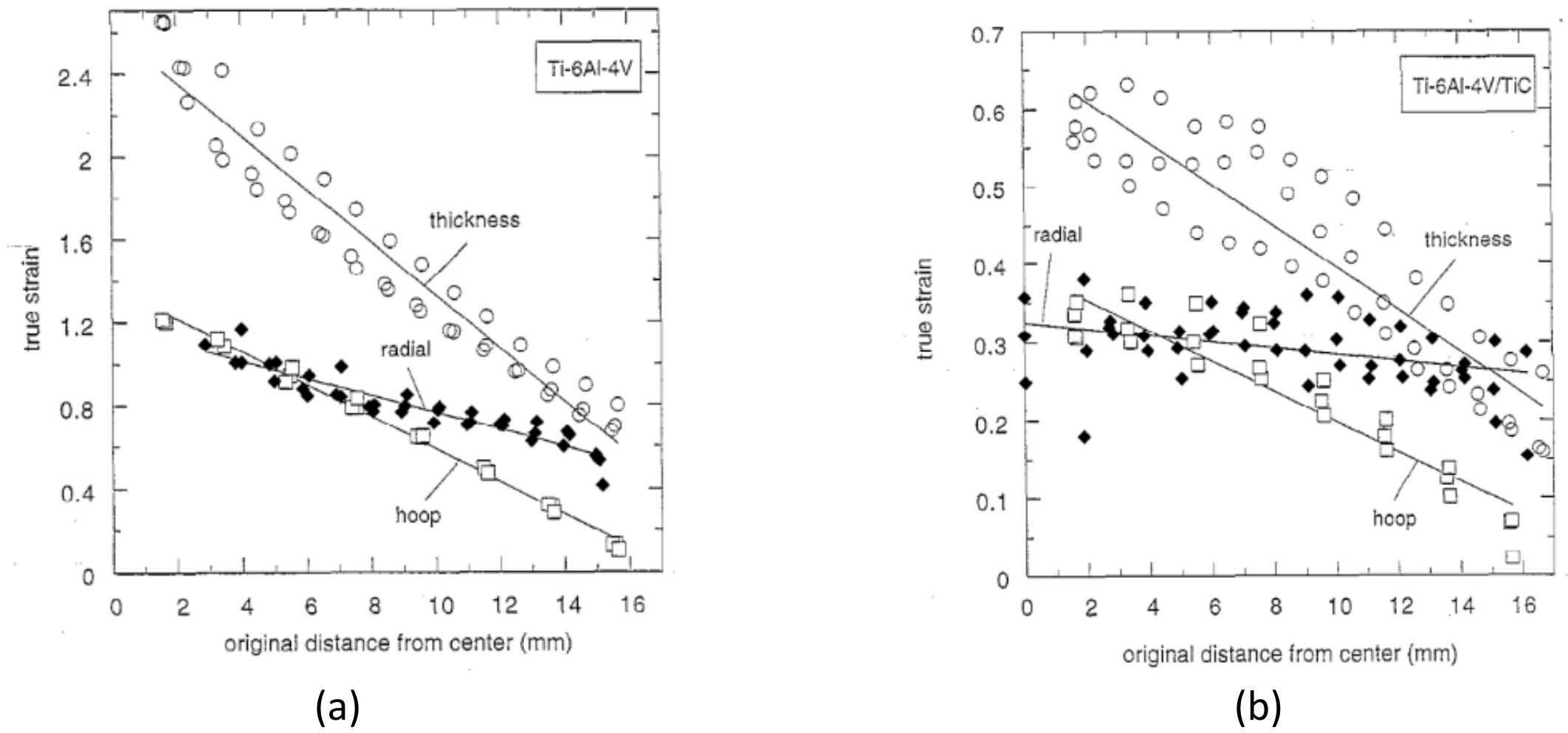

*Figure 26: Principle strains measured at different locations from the centroid for (a) Ti-6Al-4V and (b) reinforced Ti-6Al-4V/TiC. Credits: [23]*

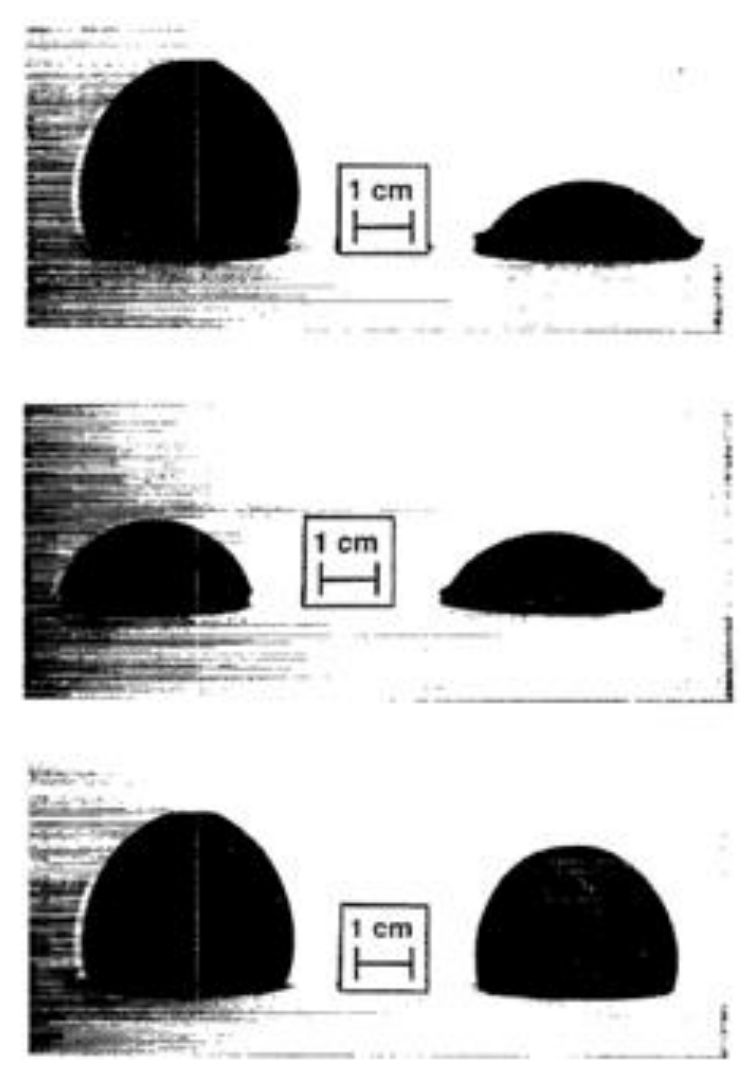

*Figure 27: Deformed domes. 1 - Ti-6Al-4V cycled (left top) and isothermal (right top); 2 – Composite Ti-6Al-4V/TiC cycled (left middle) and isothermal (right middle); 3 – Samples cycled to fracture: Ti-6Al-4V (left bottom) and Ti-6Al-4V/TiC (right bottom). Credits: [23]*

It has been found after intensive study and experimentation that microstructure of the alloy being superplastically formed has a vast impact on the SPF temperature and surface finish of the product. To understand these impacts of microstructure on SPF, the next section has been dedicated to explaining the influence of microstructure on SPF including the commonly used methods to modify the microstructure, the mechanism of SPF at the microstructure level, various characterization techniques used and results obtained by multiple research groups.

# 4.0 Microstructure

Superplastic forming of Ti-6Al-4V is a high temperature forming process for which the temperatures go up to about 1000-1200℃. After a lot of studies and experimentation it has been found that the microstructure has a huge influence on the superplasticity of the alloy. Hence, further studies have been carried out to understand the exact effect of microstructure on the same. Currently, vast amount on research is focused on modifying the microstructure to lower the superplastic forming temperatures. Not just this, but also studies are being undertaken on reducing the oxidation of titanium alloy at such high temperatures and maintaining a good surface finish along with uniform thickness of the superplastically formed component. As we further proceed into this section of the paper, we will see more about the various aspects relating to microstructure and superplasticity.

## 4.1 Methods to modify the microstructure and grain size

The grain size of the commercially procured Ti-6Al-4V alloy has to be reduced down to be submicro size to achieve relatively lower temperatures for superplastic forming. This modification is done in a different way by every researcher. A few common ways to reduce the grain size are hot rolling, equal channel angular pressing, high pressure torsion, multi axial forging and hot pressing.

### 4.1.1 Hot forging

A rod of Ti-6Al-4V alloy was procured. It was isothermally forged in a multistep process in the temperature range of 700-600℃. the forged alloy was then flat rolled at 650℃ down to a thickness of 0.8mm. To retain the submicron-sized grains and to produce a sheet with isotropic properties a low temperature range was chosen. This process led to a grain size of about 0.4µm which is submicron grain size and showed a peak elongation of 1200% at a temperature of 750℃ and a strain rate of $7 \times 10^{-3}$ $s^{-1}$. The results obtained in this method can be observed from *Table 6.*

Table 6: SPF results of hot forged Ti-6Al-4V sheet

| T (℃) | Strain rate ($s^{-1}$) | Flow stress at 50% of strain | | | Flow stress at 100% of strain | | | Elongation (%) | | | Coefficient (m) | | |
|---|---|---|---|---|---|---|---|---|---|---|---|---|---|
| | | 0° | 45° | 90° | 0° | 45° | 90° | 0° | 45° | 90° | 0° | 45° | 90° |
| 650 | $7 \times 10^{-4}$ | 70 | 71 | 71 | 74 | 74 | 74 | 780 | 830 | 810 | 0.42 | 0.45 | 0.44 |
| | $7 \times 10^{-3}$ | 179.5 | 180 | 184 | 187 | 186.5 | 190 | 720 | 660 | 680 | 0.36 | 0.32 | 0.34 |
| | $3 \times 10^{-2}$ | 315 | 317 | 309.5 | 307.5 | 310 | 297.5 | 250 | 230 | 270 | 0.22 | 0.21 | 0.24 |
| 700 | $7 \times 10^{-4}$ | 44 | 48 | 46 | 54 | 57 | 55 | 900 | 910 | 900 | 0.51 | 0.51 | 0.52 |
| | $7 \times 10^{-3}$ | 110.5 | 111 | 118 | 100 | 104.5 | 104 | 890 | 900 | 900 | 0.47 | 0.47 | 0.47 |
| | $3 \times 10^{-2}$ | 165 | 165 | 165 | 160 | 160 | 159 | 550 | 570 | 600 | 0.40 | 0.40 | 0.42 |
| 750 | $7 \times 10^{-4}$ | 27 | 26 | 29 | 37 | 36 | 40 | 1000 | 970 | 960 | 0.65 | 0.62 | 0.62 |
| | $7 \times 10^{-3}$ | 95 | 85 | 80.5 | 95 | 95 | 94 | 1200 | 1000 | 1100 | 0.51 | 0.49 | 0.51 |
| | $3 \times 10^{-2}$ | 151.5 | 146 | 146.5 | 152 | 148 | 148 | 500 | 520 | 500 | 0.41 | 0.43 | 0.41 |

### 4.1.2 High-Pressure Torsion (HPT)

Several reports show that High-Pressure Torsion is the optimum technique to produce smallest grain sizes for superplasticity of Ti-6Al-4V; nanocrystalline Ti-6Al-4V alloys have been processed using this technique. A solution of the alloy is annealed at high temperatures of about 1250K and then cooled at various rates to obtain either alpha martensitic or alpha-beta lamellar microstructure. The alloy after initial heat treatment is further processed to produce disks of thickness 8mm and diameter of 10mm by HPT at room temperature using a rotation speed of 1 rpm through a total of 20 revolutions at an applied pressure of 6 GPa under quasi-constrained conditions. Different kinds of microstructures yield different values of hardness and final microstructures (after the superplastic forming). It was observed that the alpha microstructure has smaller grains and higher hardness values than the alpha-beta lamellar microstructure. The initial microstructure also defines the volume fractions of boundaries that act as nucleation sites for rapid grain fragmentation and subgrain formation during the further forming steps. The strains observed with different microstructures at different temperatures can be seen in the table below (*Table 7).*

*Table 7: Strains in HPT sheets of Ti-6Al-4V after SF*

| **Initial Microstructure** | **Temperature (K)** | **Strain rate ($s^{-1}$)** | **δ (%)** | **Ref.** |
|---|---|---|---|---|
| α + lamellar (α + β) | 923 | $1 \times 10^{-3}$ | 575 | Sergueeva et al. [13] |
| α + lamellar (α + β) | 923 | $1 \times 10^{-3}$ | 568 | Sergueeva et al. [13] |
| | 998 | $1 \times 10^{-2}$ | 504 | |
| | 998 | $1 \times 10^{-3}$ | 676 | |
| α + 75% lamellar (α + β) | 873 | $1 \times 10^{-3}$ | 540 | Fu et al. [16] |
| | 873 | $1 \times 10^{-4}$ | 790 | |
| lamellar (α + β) | 973 | $1 \times 10^{-2}$ | 410 | This paper |
| | 973 | $1 \times 10^{-3}$ | 690 | |
| Martensite α′ | 923 | $1 \times 10^{-2}$ | 500 | This paper |
| | 973 | $1 \times 10^{-2}$ | 610 | |
| | 973 | $1 \times 10^{-3}$ | 815 | |

### **4.1.3 Rapid annealing**

A commercially available grade of Ti-6Al-4V can be rapidly annealed into beta phase region by using a direct resistance heating technique. Thermocouples are placed at the middle of the specimen to monitor the temperature. After a redefined temperature is reached, the power is switched off and the specimen is allowed to cool in atmospheric air. Rapid annealing leads to transformation of microstructure with acicular alpha grains in beta matrix. This technique doesn't vary the beta grain size by much and they remain about 10-15µm in size which is finer than usual processing techniques. The dynamic recrystallization undergone during the rapid annealing can be seen in the stress-strain curves shown in *Figures 27 and 28.*

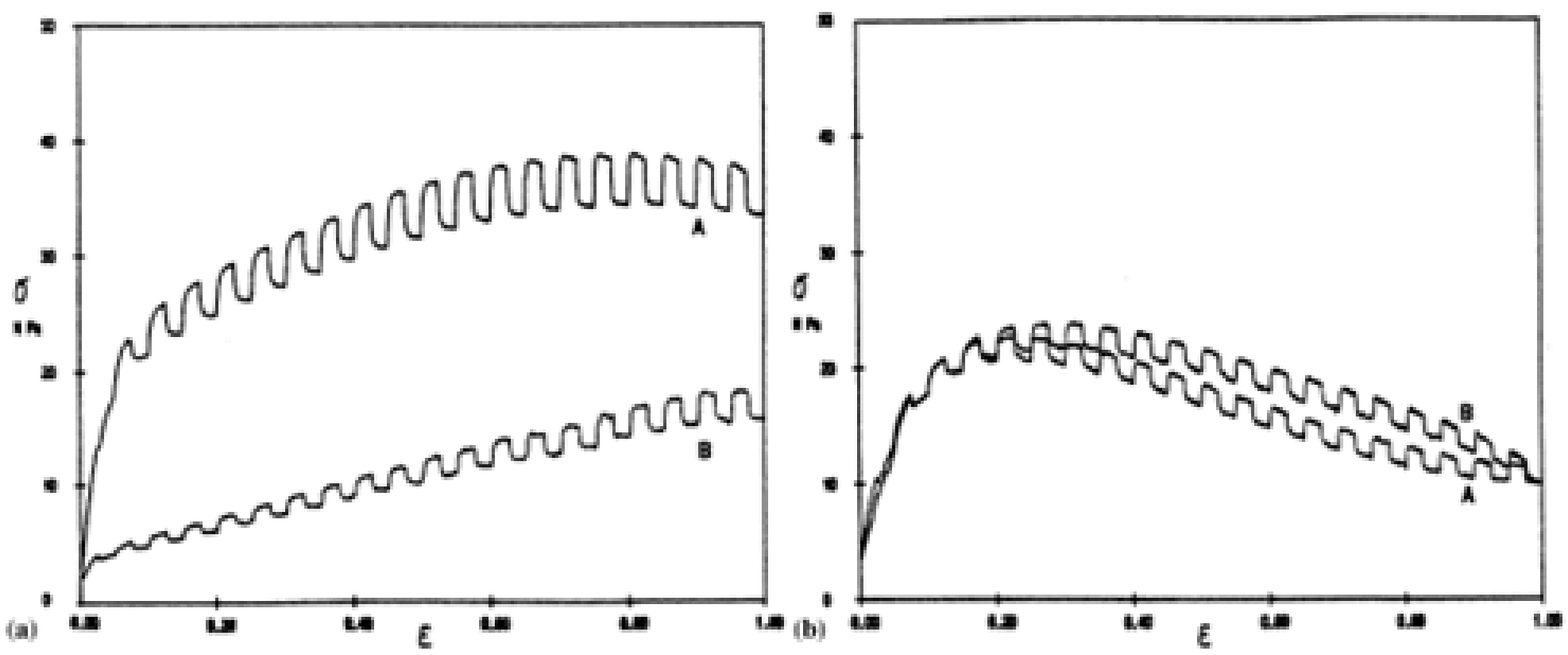

*Figure 28: Stress-strain curves a) strain rate of $10^{-3}$ $s^{-1}$ and temperature of 1025°C b) strain rate of $10^{-4}$ $s^{-1}$ and temperature of 1070°C*

## 4.2 Influence of microstructure on superplasticity

The microstructure of the alloy plays a vital role in the superplasticity of Ti-6Al-4V alloy. Studies have shown that the beta phase volume has a huge influence on the superplastic behaviour of the alloy. The beta phase of Ti-6Al-4V is the deformable part of the microstructure grains. Superplastic formation is obtained by the beta phase inhibiting grain growth and thus promoting the sliding of grain boundaries. Usually, a lamellar type of microstructure is preferred for superplastic forming due to the presence of a significant amount of beta phase unlike the case of alpha martensitic. The only drawback of having higher volume of beta phase is that it reduces the hardness and strength of the alloy, alpha phase is responsible for the above mentioned mechanical properties.

The superplastic forming process is a high temperature forming process. Hence, there is a possibility of oxidation of the surface which is undesirable for this process. Experimental results have shown that fine grain microstructures have a higher tendency to absorb oxygen than the coarse grain material. This leads to the formation of multiple oxide layers on the surface, usually above 700℃. These layers primarily consist of the oxides of aluminium and titanium and the number of layers also increase with increased exposure time and temperature. This increased level of oxygen can be associated with the increased micro hardness and reduced ductility in the alloy. The diffusion starts in the direction parallel through the alpha and beta grains and along the grain boundaries. As the concentration of increases above that needed to convert beta phase to alpha phase, a surface layer known as the alpha case is formed. This results in a brittle oxidised material. The oxygen content tends to increase with the increase in alpha titanium grains as they are known to be oxygen stabilisers. To prevent the said problem due to the interaction with oxygen, the processing is done either in inert atmosphere or coatings are used on the alloy to prevent surface interaction with oxygen. Sodium silicate and Delta Glaze 3410 are tested anti-oxidant coatings used for titanium alloys. These prevent the oxidation of the surface and also lead to a better surface finish along with better mechanical properties. The formed part with and without the glaze can be seen in *Figure 29.* After a lot of studies, it has been concluded that the elongation to failure in tensile testing generally increases with a decrease in gauge length.

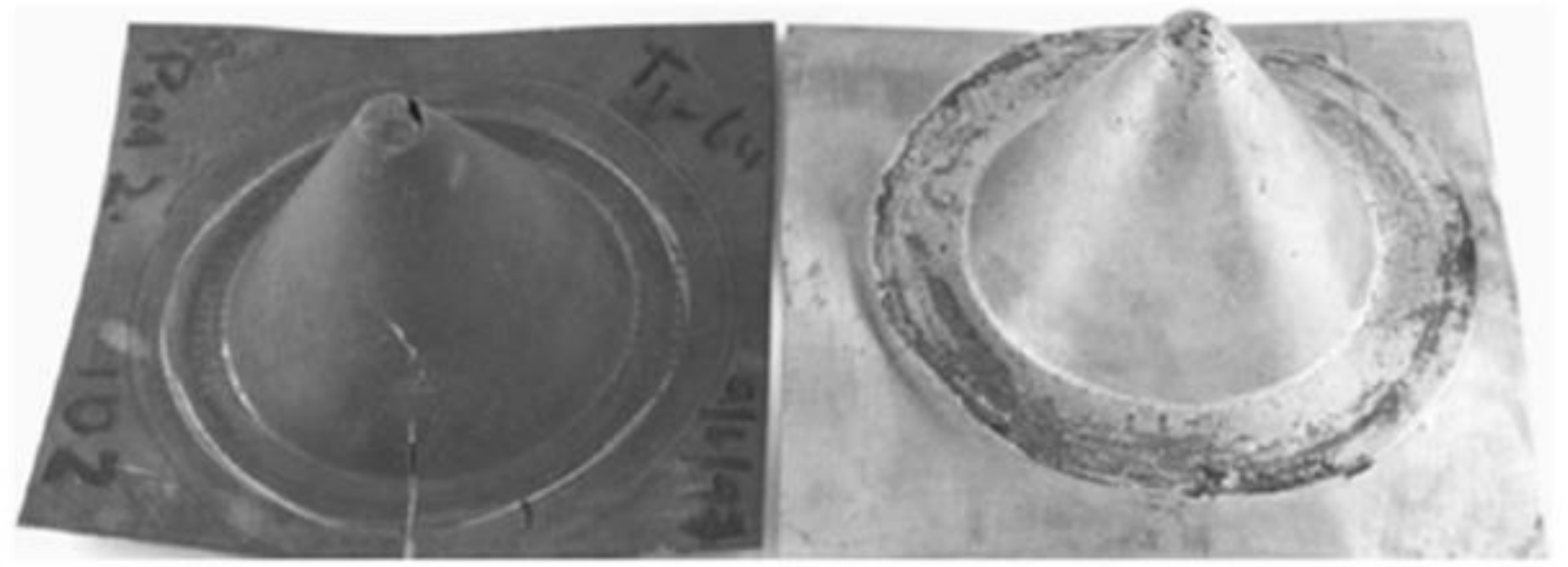

*Figure 29: Surface finish of formed part, with coating (left) and without coating (right)*

## 4.3 Mechanisms of superplasticity

The material is said to show hardening initially which is followed by a peak stress and gradual softening [3]. Significant amount of residual stress, dislocations and stacking faults along with shear transformation (as seen in *Figure 30)* and supersaturated chemical composition is observed at the end of the superplastic forming process. At higher deformations, dislocation slip and phase transformations could get activated to accommodate the deformation. One of the major transformation during the process could be that of hcp to fcc. The percentage elongation depends a lot on the grain boundary sliding that can occur in the particular sample taken for superplastic forming. At very low strain rates negligible dynamic coarsening of the microstructure is observed. The beta-phase precipitation after holding at the testing temperature acts as an accommodating mechanism and enhances superplasticity; even a small amount of beta-phase contributes to superplasticity by inhibiting grain growth and promoting grain boundary sliding. From this, it can be assumed that the volume fraction of beta phase determines the dominant flow mechanism. The superplastic flow also depends on the crystal lattice as various fcc and hcp have different number of slip systems as they help inhibit grain growth and promote grain boundary sliding (fcc has higher number of slip systems and undergoes intragranular slip to accommodate grain boundary sliding).

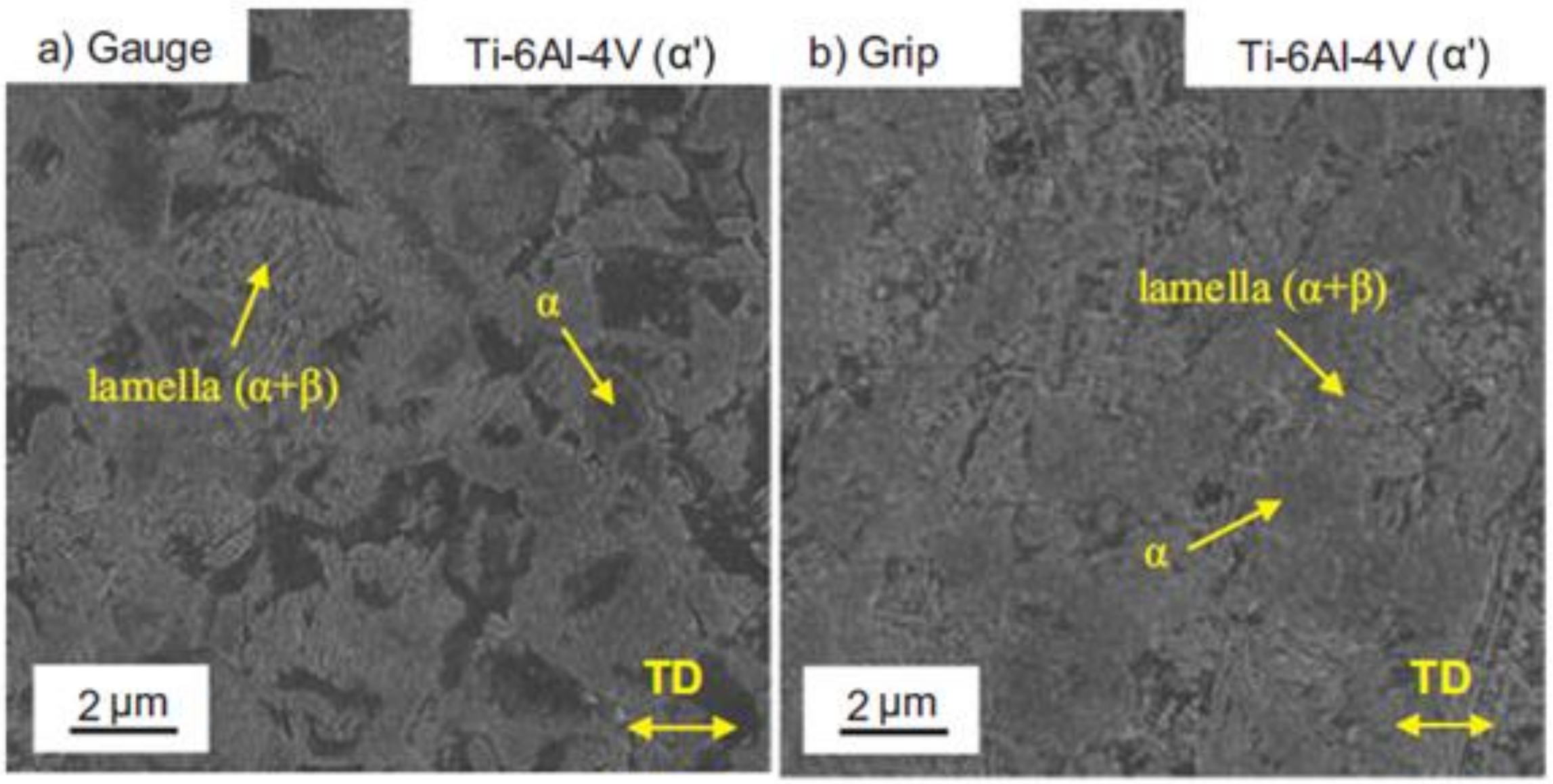

*Figure 30: Microstructure of the tensile test sample observed under an optical microscope a) at gauge b) at grip*

The microstructural evolution takes place at different rates across the specimen until it becomes almost homogenous [5]. The initial processing done for microstructural refinement produces a high density of dislocations and dislocation tangles. Here, the microstructural flow and presence of shear bands are revealed using selected area electron diffraction (SAED). With increasing shear in the pre-processing stages the SAED patterns become more ring-like and thus showing grain misorientations and evolution of high angle grain boundaries. During the superplastic forming, strain hardening is observed at higher strain rates and superplastic flow with low stable flow stresses are observed at low strain rates. From the limited set of data, it has been observed that the strain hardening constant (m) increases and then decreases with increasing strain rate (*Figure 31)*. It has also been observed that grain boundary sliding in superplasticity requires m = 0.5. Common to all the studies, an alpha to beta phase transformation takes place in the alloy during prolonged exposure at the tensile deformation temperature (>750℃). It has also been concluded that high angle grain boundaries are crucial for the flow via extensive grain boundary sliding.

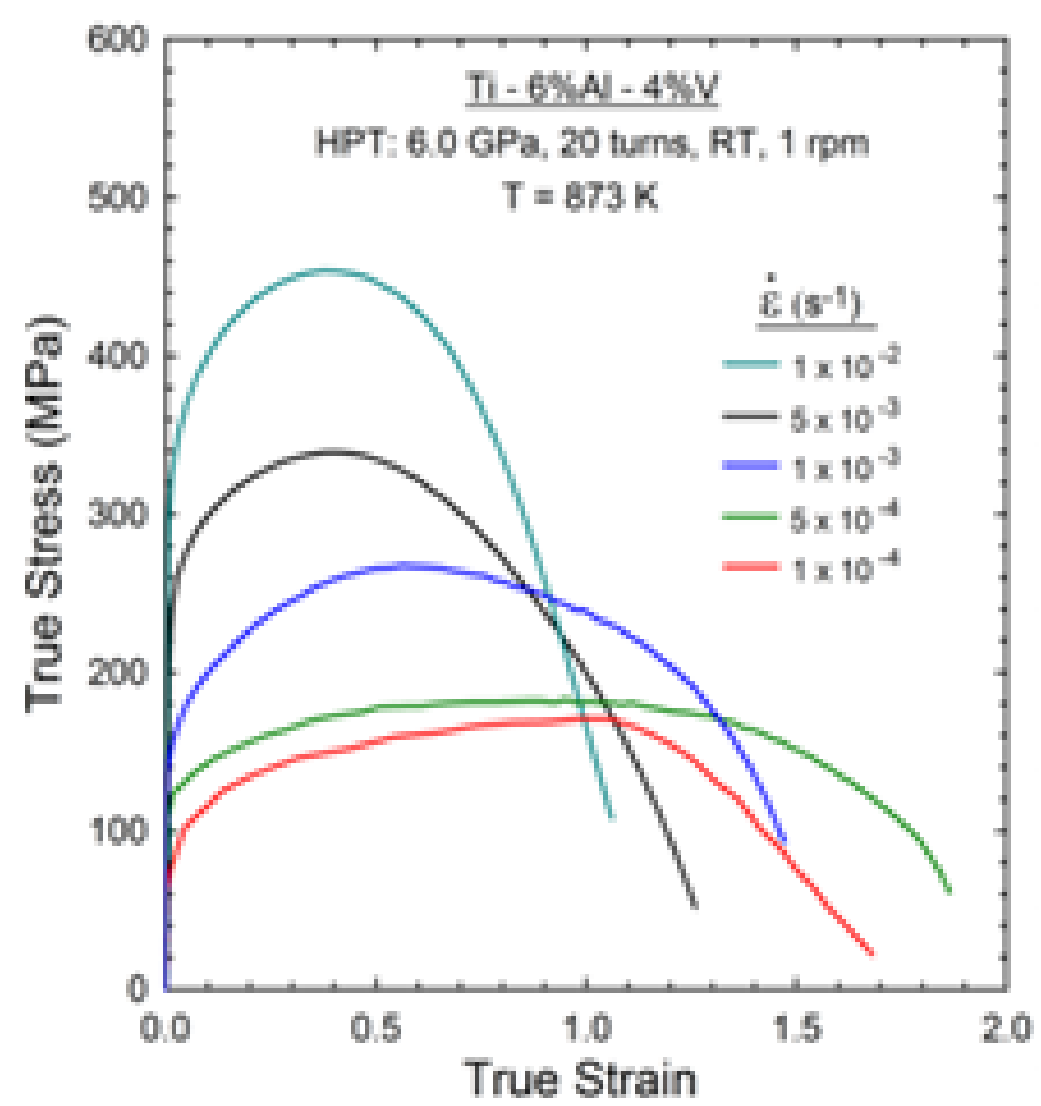

*Figure 31: Stress-strain curves at varied strain rates*

The superplastic deformation is a result of dislocation gliding, climbing and grain boundary sliding [1]. At the superplastic forming temperature (below 800℃), the mechanism is dominated by accumulation of dislocations as the thermal activation of grain boundary sliding gets difficult. This increase in dislocation density leads to dislocation tangling and subgrain formation. At a relatively low temperature, dynamic recovery can act as an accommodating mechanism. The dislocations at this temperature could be of edge dislocations, screw dislocations, de-pinning of dislocation or counteraction of dislocation to cancel each other out. For the temperature above 800℃, superplastic forming occurs through grain boundary sliding accompanied by dynamic recrystallization. This is due to the matching of the activation of energy of superplastic forming with the grain boundary self-diffusion energy. Most of the dislocations are observed at alpha-beta phase boundary and not the alpha-alpha phase boundary. Thus, it can be concluded that the superplastic forming of two-phase Ti-6Al-4V vastly depends on the alpha-beta phase ratio. Moreover, beta phase is soft and acts as a lubricant during the deformation providing stress relaxation. The above mentioned microstructural changes can be observed in the TEM images shown in *Figure 31.*

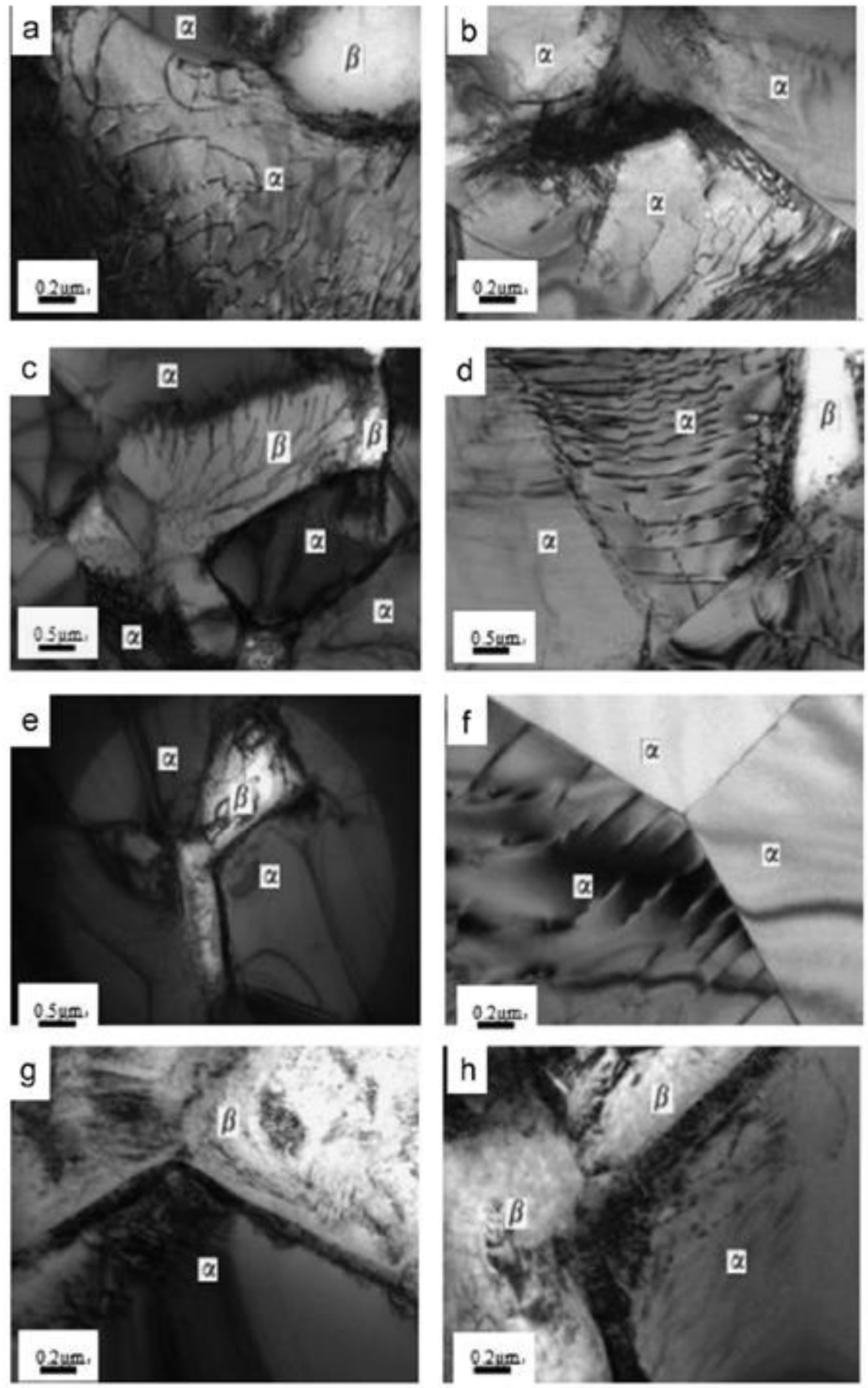

*Figure 32: TEM images of tensile test samples (a & b) - 700°C, (c & d) - 750°C, (e & f) - 800°C, (g & h) - 850°C*

After years of research it has been concluded that very fine grain size and high temperature are the required conditions to achieve superplastic behaviour (Alabort, Putman & Reed, 2015). But these are two that work in the opposite directions due to the grain growth observed at elevated temperatures.

Most of the researchers have concluded that grain boundary sliding is mechanism by which superplastic forming can be attained, but is still unclear as to if the grain boundary sliding is diffusion-accommodated, dislocation-accommodated or combinations of both. There also could be cavity formation as the part of the mechanism for superplasticity. Hence, it is fair to assume that there are multiple mechanisms behind superplastic forming. The only certain thing is that two-phase microstructure is critical for superplastic forming. These effects of microstructural changes on the stress-strain plot can be seen in *Figure 33.*

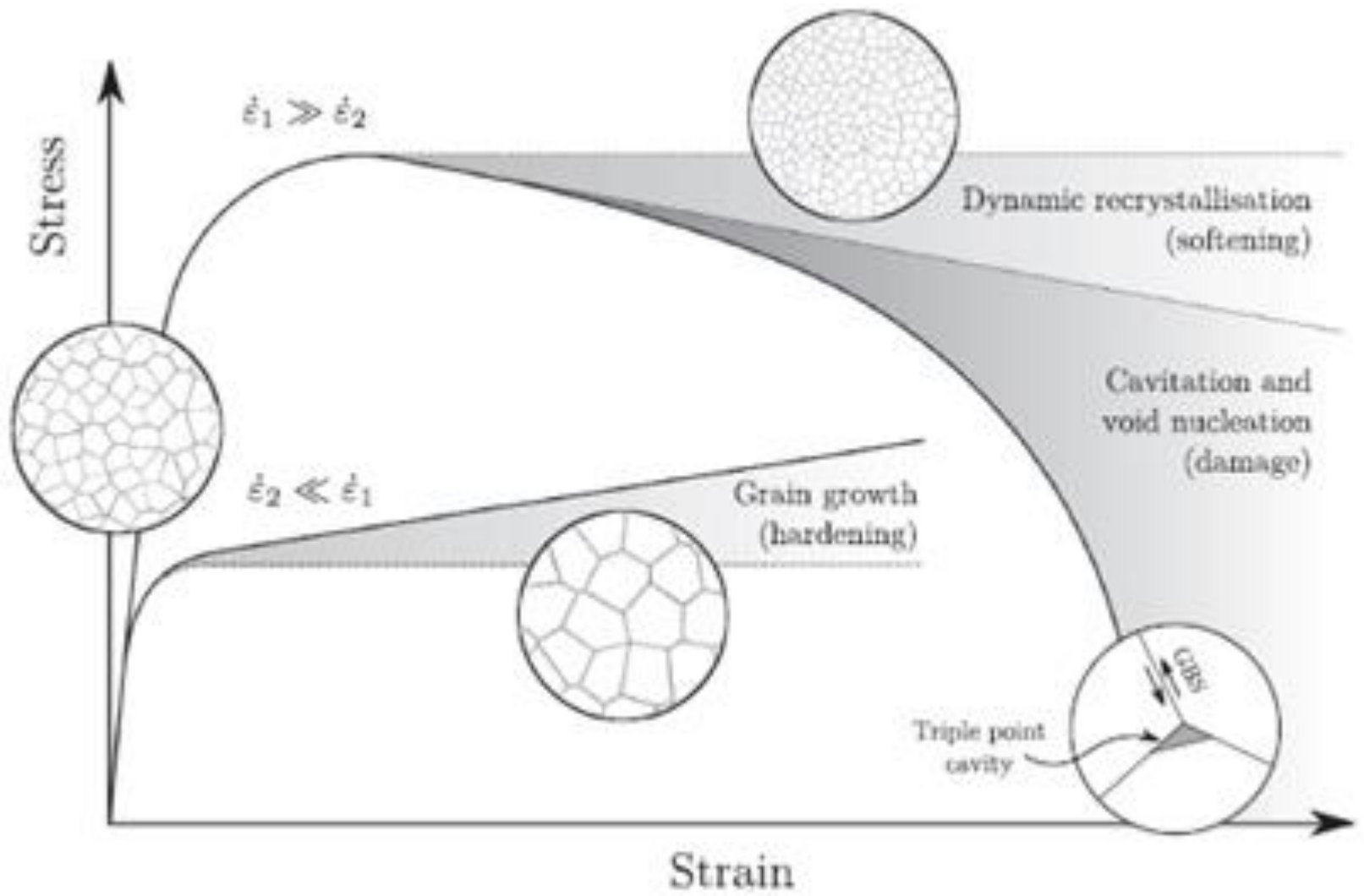

*Figure 33: Effects of microstructural changes on the stress-strain curve*

## 4.4 Microstructure study

| **S.No.** | **Technique used** | **References** |
|---|---|---|
| 1 | Optical Microscopy | 3; 10; 11; 12; 13; 14; 15; 17; 18; 19; 20; |
| 2 | Scanning Electron Microscopy | 1; 10; 12; 16; 18; |
| 3 | Transmission Electron Microscopy | 1; 2; 3; 5; 9; 13; |
| 4 | Orientation Imaging Microscopy | 1; 8; 19; |
| 5 | Vicker’s Microhardness | 3; |
| 6 | Electron Back Scattered Diffraction | 1; 3; 5; 8; 10; 12; 17; 18; 19; |
| 7 | X-Ray Diffraction | 2; 3; 5; 17; |
| 8 | Thermal Field Emission Scanning Electron Microscope | 3; |
| 9 | Selected Area Electron Diffraction | 5; |
| 10 | Inductively Coupled Plasma Spectroscopy | 6; |
| 11 | Knoop Hardness | 6; |
| 12 | Ultrasonic C-Scan | 18; |

## 4.5 Experimental results

| **S.No.** | **Grain size** | **Temperature (°C)** | **Strain rate ($s^{-1}$)** | **Max. Elongation (%)** | **Reference** |
|---|---|---|---|---|---|
| 1 | 2.5µm | 800 | $1x10^{-3}$ | 862 | 1 |
| 2 | 0.4µm | 750 | $7x10^{-3}$ | 1200 | 2 |
| 3 | 40nm | 700 | $1x10^{-3}$ | 815 | 3 |
| 4 | 1µm | 775 | | | 4 |
| 5 | 40nm | 650 | $5x10^{-4}$ | 820 | 5 |
| 6 | 3µm | 775 | $1x10^{-4}$ | 850 | 8 |
| 7 | 0.5µm | 800 | $1x10^{-3}$ | 700 | 11 |

| 8 | | 649 | | 37 | 18 |
|---|---|---|---|---|---|
| 9 | 10µm | 800, 850, 900, 950 | 1x$10^{-4}$ | >300 | 19 |

The work on superplastic forming of Ti-6Al-4V alloy has been carried out by many around the world. A few of the works on SPF/DB have been patented by the creators. A peek into those patents has been provided in the section below.

## 5.0 Patents on SPF/DB

A study of SPF/DF found that a combination of superplasticity and diffusion bonding is performed at low pressures [31]. An innovative approach is described which uses sub-atmospheric pressures between the to-be formed sheets instead of externally applied force through hydraulic press. For fine grained Ti-6Al-4V (grain size less than 2 micrometer), an approach to conduct SPF/DB at a reduced temperature and increased strain rate thereby decreasing the cost of the process [32].

The work on SPF and its applications does not end here. This is a field with a lot of potential and scope for the future. This is due to its nature which can replace the conventional forming processes once the technique has been optimised and perfected. The work on SPF has been going on for about 30 years now and there have been drastic developments in the methods in these three decades. After, a careful study of a lot of research work in this field, a few potential areas of advancements and applications have been brought out in the next section. This will help uplift the further research and developments in the said field.

## 6.0 Scope

For future work, a combination of higher applied stresses and faster temperature cycling in the case of mix-match SPF may allow for increase in the average strain rate by about an order of 10 [23]. This implies that it would be possible to achieve higher deformation rates similar to those achievable in optimized Ti-6Al-4V materials having microstructural superplasticity. Thus, transformation mix-match SPF could become a well-sought-after forming method for Titanium alloys which are difficult to forge, machine or cast.

The superplasticity of Ti-6Al-4V alloy is improved by additions of traces of hydrogen which decrease the dislocation density, promote dislocation motion and facilitate Beta phase flow [13].

Rapid Beta heat treatment and the microstructural gradient technique can be implemented in a controlled way to manipulate the local thickness straining [7]. Their feasibility will be improved if the problem at the interface zone has been well addressed, as well as better prediction and control of applying the gradients, even though it may be difficult in practice. There is also a remote possibility that, if a precise prediction and control of the microstructural gradients can be achieved, highly complex shapes might be even formed by free-bulging without a die, i.e. die-less forming.

From an experimental viewpoint, multi-step dies lead to wrinkling, which although removable through "ironing" in the second step, leads to decrease in the structural strength of the alloy [33]. In addition, adopting a multi-stage die with a gradient in coefficient of friction results in more uniformity when compared to single-step dies [28]. Further research into FEA-guided manufacturing of multi-step dies is required so as to optimise the design such that no wrinkles are present in the final die. Experimental conditions given in indicate that the optimal conditions for SPF are 900-980℃ [30] and 5-10 MPa [29] with strain rates between 0.001/s and 0.0001/s [19]. Moreover, optimal pressure conditions decrease coefficient between the die and alloy [27]. The optimal pressure and temperature curves are somewhat inverted bell-shaped. Although, increasing the SPF pressure optimises the process, indefinitely increasing temperature and pressure leads to loss in structural integrity causing deformations and wrinkles in the structure. Forging parameters need to optimised as per the above discussion, which is possible via microstructure refinements and deeper understanding of grain formation during SPF. Further research in to achieving sub-micro grained Ti-6Al-4V should be done as shown by [8] that such Ti-6Al-4V showed more deformation, hence, more elongation as compared to traditional Ti-6Al-4V. Furthermore, [34] showed that sub-microcrystalline (SMC) structured Ti-4Al-4V underwent a relatively more uniform cross-sectional strain than commercial alloys with microcrystalline (MC) structure. However, SPF temperature for commercial grade Ti-6Al-4V is spread over a wider range (700-1100℃) [35] and occurs at a lower strain rate than that mentioned in [29]. Therefore, further research in obtaining desired physical properties of SPF, as in [2], at lower temperatures (700℃) need to be investigated. For instance, as [9] and [1] found, nanometric grains and narrow grain distribution lead to lowering of optimum SPF temperature. Recrystallisation is the main microstructural process which occurs throughout the SPF [17]. Hence, further studies to identify the factors which lead to optimal recrystallisation is required. Furthermore, quantitative models to predict thickness distribution based on initial microstructure (along with its evolution through the process) need to be built to correlate grain size and phase concentrations for a generalization of the constitutive behavior of the SPF process [37]. Also, [6] extensively showed that fine-grained Ti-6Al-4V showed greater oxygen diffusion rates in the areas of α-case thickness and microhardness depth as compared to the commercial counterpart. However, this also means that present manufacturing techniques need to be revised to account for this increased diffusion while accurately studying the effects of this oxygen enrichment on the physical properties of the alloy.

Microstructural reasons of high SPF are given as suppression of porosity and high stability of the grain formation regions [38]. Therefore, research into replicating these properties in desired metals could be an interesting avenue for future work, leading to commercial SPF of other Titanium alloys. Other major factors which influence coefficient of friction and strain rate wherein low strain rate resulted in more uniform thickness distribution of the formed alloy [39]. Thickness nonuniformity, as shown by [17], leads to stress concentrations causing localized thinning and structural integrity degradation. Hence, a theory specifying selective heating areas to avoid such nonuniformity based on the geometry of the forming setup is required.

## 7.0 Conclusions